\documentclass[10pt,twocolumn,journal]{IEEEtran}
\usepackage{latexsym}
\usepackage{amsfonts}
\usepackage{amsbsy}
\usepackage{amsmath,amssymb}
\usepackage{times}
\usepackage{graphicx}
\usepackage{enumerate}
\usepackage[usenames]{color}
\usepackage[dvips]{pstcol}
\usepackage{epstopdf}
\usepackage{cite}
\usepackage{amsmath}
\usepackage{amssymb}
\usepackage{amsfonts}
\usepackage{graphicx}
\usepackage{epsfig}
\usepackage{psfrag}
\usepackage{xcolor}
\usepackage{amsfonts, bm}
\usepackage{epstopdf}
\usepackage{cite}
\usepackage{color}
\usepackage{xcolor}
\usepackage{subfig}
\usepackage{booktabs}

\newcommand{\st}{{\rm s.t.}}

\newcommand{\bF}{\mathbf{F}}

\input epsf

\usepackage{lineno}
\usepackage{algorithm} 
\usepackage{multirow} 
\usepackage{algpseudocode}
\usepackage{graphics}
\usepackage{epsfig}
\usepackage{subeqnarray}
\usepackage{cases}
\usepackage{setspace}
\usepackage{diagbox}
\usepackage{graphicx}
\usepackage{subfig}
\usepackage{subfloat}
\pdfoutput=1

\begin{document}
\title{  Deep-Unfolding Neural-Network Aided Hybrid Beamforming Based on Symbol-Error Probability Minimization}


\author{Shuhan Shi, Yunlong Cai,  Qiyu Hu, Benoit Champagne, and Lajos Hanzo
\thanks{
S. Shi, Y. Cai, and Q. Hu are with the College of Information Science and Electronic Engineering, Zhejiang University, China (e-mail: ssh16@zju.edu.cn; ylcai@zju.edu.cn; qiyhu@zju.edu.cn).
B. Champagne is with the Department of Electrical and Computer Engineering, McGill University, Canada (e-mail: benoit.champagne@mcgill.ca).
L. Hanzo is with the Department of Electronics and Computer Science, University of Southampton, U.K. (e-mail: lh@ecs.soton.ac.uk).
}
}

\maketitle
\begin{abstract}
		In massive multiple-input multiple-output (MIMO)
                systems, hybrid analog-digital (AD) beamforming can be
                used to attain a high directional gain without
                requiring a dedicated radio frequency (RF) chain for
                each antenna element, which substantially reduces both
                the hardware costs and power consumption. While
                massive MIMO transceiver design typically relies on
                the conventional mean-square error (MSE) criterion,
                directly minimizing the symbol error rate (SER) can
                lead to a superior performance.  In this paper, we
                first mathematically formulate the problem of hybrid
                transceiver design under the minimum SER (MSER)
                optimization criterion and then develop a MSER-based
                gradient descent (GD) iterative algorithm to find the
                related stationary points. We then propose a
                deep-unfolding neural network (NN), in which the
                iterative GD algorithm is unfolded into a multi-layer
                structure wherein a set of trainable parameters are
                introduced for accelerating the convergence and
                enhancing the overall system performance. To implement
                the training stage, the relationship between the
                gradients of adjacent layers is derived based on the
                generalized chain rule (GCR). The deep-unfolding NN is
                developed for both quadrature phase shift keying
                (QPSK) and for $M$-ary quadrature amplitude modulated
                (QAM) signals and its convergence is investigated
                theoretically. Furthermore, we analyze the transfer
                capability, computational complexity, and
                generalization capability of the proposed
                deep-unfolding NN. Our simulation results show that
                the latter significantly outperforms its conventional
                counterpart at a reduced complexity.
\end{abstract}



\vspace{-2mm}
\section{Introduction}
\label{sec:intro}

The massive multiple-input multiple-output (MIMO) technology has
inspired wide research-attention, since it is capable of dramatically
increasing the system capacity, hence mitigating the spectrum shortage
\cite{mmWave1, mmWave2, mmWave3_hybrid4}. This is achieved by forming
directional beams. However, these large-scale antenna arrays employed
both at the transmitter and receiver have a prohibitive hardware cost
and power consumption in fully-digital (FD) beamforming, where each
antenna element requires a dedicated radio frequency (RF) chain. Thus,
hybrid analog-digital (AD) beamforming has become an important
technique of reducing the number of RF chains, while still approaching the
performance of fully-digital beamforming \cite{mmWave3_hybrid4, hybrid1,
  hybrid2, 2013Spatially, 2016Hybrid, 2015MSE, 2016DH, 2017DH,
  2016Alternating, codebook, 2017GEVD, 2019GEVD}.

\vspace{-2mm}
\subsection{Prior Art}
Hybrid AD beamforming conceptually relies on the decomposition of the
beamforming operation as the product of a low-dimensional digital
beamforming matrix and a high-dimensional analog beamforming matrix.
The elements of the analog beamforming matrix obey the unit-modulus
constraint imposed by the phase shifters.  To reduce the number of RF
chains, several authors~\cite{hybrid1,hybrid2} advocated
antenna-selection based hybrid beamforming. In order to further
improve their performance, spatially sparse precoding techniques were
developed by exploiting the millimeter-wave (mmWave) channel
characteristics in \cite{mmWave3_hybrid4,2013Spatially}.  Moreover,
the authors of~\cite{2016Hybrid} demonstrated that hybrid AD beamforming
having twice as many RF chains as data streams approaches the
performance of FD beamforming.

Inspired by these findings, sophisticated optimization algorithms were
proposed for hybrid beamforming~\cite{2013Spatially, 2015MSE, 2016DH,
  2017DH, 2016Alternating, codebook, 2017GEVD, 2019GEVD}. In
\cite{2013Spatially}, the authors leveraged the channel's sparsity and
proposed an orthogonal matching pursuit (OMP) aided algorithm. Based
on the OMP concept, the authors of ~\cite{2015MSE, 2016DH, 2017DH}
developed a hybrid transceiver architecture relying on the minimum
mean square error (MMSE) criterion. In~\cite{2016Alternating}, the
authors formulated a matrix factorization problem and developed a
manifold optimization (MO) based hybrid beamforming algorithm, where
the unit-modulus constraint was considered as a Riemannian manifold.
In~\cite{codebook}, a codebook-based hybrid beamforming design was
conceived for maximizing the system's spectral efficiency. As a
further development, the authors of~\cite{2017GEVD} conceived a
two-stage optimization algorithm based on the general
eigen-decomposition method and their work evolved further
in~\cite{2019GEVD} to the broadband scenario with the aid of MO.

However, most of the existing hybrid beamforming designs in the open
literature are based on the minimization of the mean square error
(MSE), which is not the most appropriate metric from a performance viewpoint in digital communications where
the standard measure is the symbol-error-rate (SER).
Recently, a number of powerful
beamformers have been designed based on the minimum SER (MSER)
criterion~\cite{MSER1, MSER2, MSER3, MSER4, MSER5}. The authors of
\cite{MSER1} and \cite{MSER2} developed a MSER-based adaptive reduced-rank
receive beamformer for enhancing the performance.  In~\cite{MSER3}, a
single-bit direct MSER-optimization based precoder was proposed for
simplifying the RF chains.  The authors of~\cite{MSER4} designed an
interference-aided precoder for minimizing the SER of the worst-case
user in the system.  As a further development, in~\cite{MSER5}, the
authors proposed a MSER-based precoder for $K$-pair MIMO interference
channels by utilizing improper signaling.  However, due to the
associated complex high-dimensional matrix inversions and
decompositions, the existing optimization algorithms suffer from high
complexity in practical implementation.

To improve the performance at a reduced complexity, researchers have
recently turned their attention to machine learning techniques for
solving a variety of problems, such as channel
decoding~\cite{DLchanneldecoding}, end-to-end
communication~\cite{DLendtoend, CNNLi}, and channel
estimation~\cite{DLchannelestimation}.  Well-trained neural networks
(NNs) are capable of learning the mapping between the system inputs
and outputs~\cite{MLP, CNN, DNN, DLhybrid1, DLhybrid2, DLhybrid3}. One
of the early attempts along this avenue appeared in~\cite{MLP}, where
a NN was trained for performing power allocation based on an iterative
weighted MMSE (WMMSE) algorithm.  The authors of~\cite{CNN} proposed a
two-stage training mechanism for maximizing the sum-rate by applying
convolutional neural networks (CNN), and further improved the system
performance through unsupervised learning.  In~\cite{DNN}, the authors
adopted an ensemble of fully connected deep neural networks (DNNs) for
optimizing the transmit power allocation.  In~\cite{DLhybrid1}, the
problem of joint antenna selection and hybrid beamforming was first
formulated as a classification problem, and then solved by a deep CNN.
Deep learning was also utilized for jointly optimizing the compressive
channel and hybrid beamforming matrices in~\cite{DLhybrid2}. In
\cite{DLhybrid3}, the authors designed a novel NN inspired by
GoogleNet, which used parallel complex convolutional blocks for hybrid
beamforming.

In the above contributions, NNs are generally treated as black-boxes,
which does not guarantee optimal performance and leads to limited
interpretability, along with limited control.  Furthermore, in massive
MIMO systems, the training overhead of these methods is expensive due
to the multi-dimensional training samples and long training time. To
overcome these issues, a model-driven network, which is referred to as
deep-unfolding, has been proposed in~\cite{DU_first}.  This method
generally unfolds iterative algorithms into multi-layer NN structures
and introduces a set of trainable parameters to accelerate convergence
and to increase the system performance.  The authors
of~\cite{DUdata1} and \cite{DUdata2} invoked this approach to unfold the gradient
descent (GD) algorithm for data detection.
In~\cite{DUchannel1,DUchannel2} the authors developed a deep-unfolding
method inspired by the normalized min-sum algorithm for the decoding
of polar codes. In~\cite{DUsparse}, a scheme based on deep-unfolding
was put forward for sparse signal recovery.  In \cite{DUprecoder1},
the authors employed the deep-unfolding method for automatically
optimizing the precoders relying on single-bit digital-to-analog
converters (DACs). The authors of \cite{DUprecoder2} adopted a
deep-unfolding NN based on the conjugate GD algorithm for constant
envelope precoding.  A finite-alphabet precoder was developed in
\cite{DUprecoder3}, which unfolded the iterative discrete estimation
algorithm into a NN.  In \cite{IAIDNN}, the authors derived the
generalized chain rule (GCR) in matrix form and proposed a
deep-unfolding NN based on the WMMSE precoding design algorithm.

\vspace{-2mm}
\subsection{Main Contributions}
\vspace{1mm}

	The MSER-based hybrid beamforming design problem in massive MIMO systems is very challenging due to the highly nonconvex objective function and unit-modulus constraints.
While the GD algorithm is a common technique that can be adopted for finding the stationary points, it requires a long time to converge and a large number of iterations.
Since the deep-unfolding method makes full use of the structure of the iterative algorithm
and only replaces some complex operations with trainable parameters, it can significantly reduce the computational complexity with only a slight performance loss.
To the best of our knowledge however, the deep-unfolding method has not been investigated in the context of hybrid beamforming based on the MSER criterion.  Hence we first mathematically formulate the problem of hybrid AD transceiver design under the direct MSER criterion in a hardware-efficient massive MIMO system.  We then develop a MSER-based GD iterative algorithm for finding the stationary points.
Specifically, we propose a deep-unfolding NN, in which the iterative
GD algorithm is unfolded into a multi-layer structure and then a set
of trainable parameters are introduced into the forward propagation
(FP) for accelerating the convergence and enhancing the system
performance.  In the training stage, the relationship between the
gradients of adjacent layers is derived based on the GCR in the back
propagation (BP). Then a deep-unfolding NN is developed for both
quadrature phase shift keying (QPSK) and $M$-ary quadrature amplitude
modulation (QAM) signals.
	While deep-unfolding NNs often lack solid theoretical support, we herein provide a detailed theoretical convergence analysis of the proposed schemes.
As a benchmark, we develop a deep learning
aided black-box based CNN for hybrid transceiver design. Furthermore,
we analyze the transfer ability, computational complexity and generalization capability of the proposed deep-unfolding NN.
Our simulation results show that the proposed deep-unfolding NN significantly outperforms the conventional algorithms and approaches the performance of the MSER-based GD iterative algorithm at a much reduced complexity.  The main contributions of this work are summarized as follows:
 \begin{itemize}
\item
We formulate the joint hybrid AD transceiver design problem based on
the MSER criterion in a massive MIMO system context, which is very
challenging to tackle due to the highly nonconvex objective function
and unit-modulus constraints imposed on the entries of the analog
beamforming matrix.  Then an unconstrained MSER-based GD iterative
algorithm is developed for finding the stationary points through the
application of kernel density estimation and a parametric
representation of the unit-modulus matrix entries.
\item
We propose a deep-unfolding NN based on the MSER-based GD iterative
algorithm to expedite convergence while guaranteeing the performance
of our hybrid transceiver design, wherein the iterative algorithm is
unfolded into a multi-layer structure and a set of trainable
parameters are introduced.
\item
	We provide a theoretical analysis for the convergence of the proposed deep-unfolding NNs and also analyze their transfer ability, computational complexity and generalization capability.
\item
	Through numerical simulation, we show that the
proposed deep-unfolding NN substantially outperforms the conventional
black-box method and approaches the performance of the MSER-based GD
algorithm at a significantly reduced number of iterations, which
translates into reduced complexity.
\end{itemize}

\vspace{-5mm}
\subsection{Organization}
\vspace{-1mm}
The paper is structured as follows. Section \ref{Section2:system}
briefly describes the system model. Section
\ref{Section3:gradient-descent} formulates the MSER-based hybrid
transceiver design problem for the case of QPSK and develops a
MSER-based GD algorithm. The deep-unfolding NN is conceived in Section
\ref{Section4:deep-unfolding}. In Section \ref{Section5:analysis}, a
black-box NN as a benchmark, the analysis of the computational
complexity and generalization ability and the extension of the
proposed algorithm to the QAM modulation are provided. The simulation
results are presented in Section \ref{Section7:simulations} and
conclusions are drawn in Section \ref{Section8:conclusion}.

\emph{Notations:} Scalars, vectors and matrices are respectively denoted by lower case, boldface lower case and boldface upper case letters.
$\mathbf{I}$ represents an identity matrix and $\mathbf{0}$
denotes an all-zero  matrix.
For a  matrix $\mathbf{A}$, ${{\bf{A}}^T}$, $\mathbf{A}^*$, ${{\bf{A}}^H}$, $\mathbf{A}^{\perp}$,  $\|\mathbf{A}\|$ and $[\mathbf{A}]_{m,n}$  denote its   transpose, conjugate, conjugate transpose, reciprocal by element, Frobenius norm and element at the
intersection of row $m$ and column $n$, respectively.
For a vector $\mathbf{a}$, $\|\mathbf{a}\|$ represents its Euclidean norm.
$\mathbb{E}\{.\}$ denotes the statistical expectation.
$\Re\{.\}$ ($\Im\{.\}$) denotes
the real  (imaginary) part of a variable.
The operator $\textrm{vec}( \cdot )$ stacks the columns of a matrix in one long column vector.
 $|  \cdot  |$ denotes  the absolute value of a complex scalar.
${\mathbb{C}^{m \times n}}\;({\mathbb{R}^{m \times n}})$ denotes the space of ${m \times n}$ complex (real) matrices.
The operator $\angle$ takes the phase angles of the elements in a matrix. The symbol $\circ$ denotes the Hadamard product of
two vectors/matrices.

\vspace{-2mm}
\section{ System Model }
\label{Section2:system}
\begin{figure}[t]
	\centering
	\includegraphics[width=0.5\textwidth]{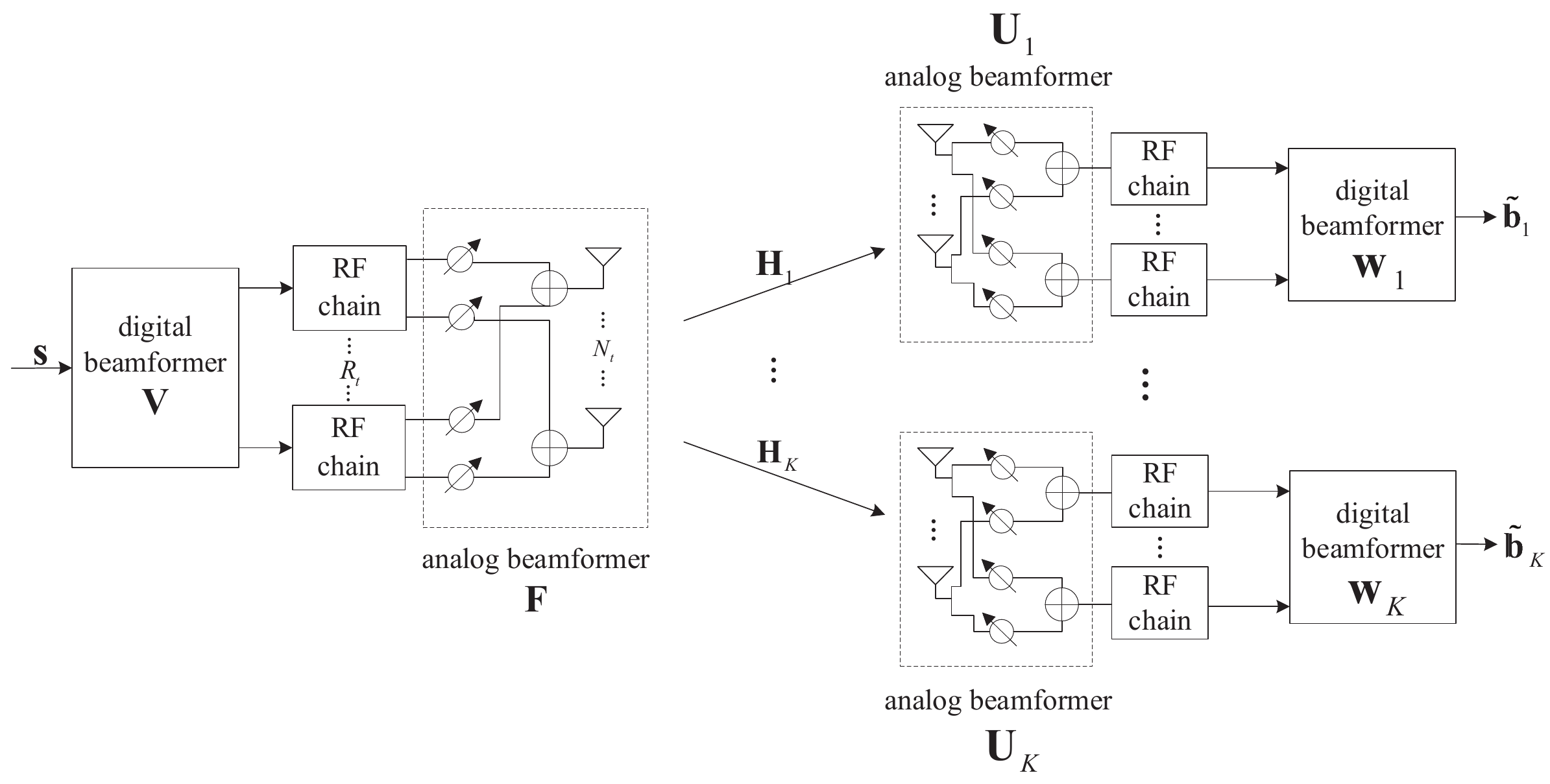}
	\caption{Hardware-efficient massive MIMO  system for downlink transmission.}
	\label{fig:structure}
\end{figure}

We consider the downlink of a hardware-efficient massive MIMO system
as depicted in Fig. \ref{fig:structure}, which consists of one base
station (BS) and $K$ users. The BS is equipped with $N_t$ transmit
antennas and $R_{t} (R_{t}\ll N_t)$ RF chains.  User
$k\in\mathcal{K}\triangleq \{1, \ldots, K\}$ is equipped with
$N_{r,k}$ receive antennas and $R_{r,k} (R_{r,k}\leq N_{r,k})$ RF
chains, where $N_{t}\gg \sum^{K}_{k=1}N_{r,k}$ and $R_{t}\geq
\sum^{K}_{k=1}R_{r,k}$.  The BS transmits a symbol vector
$\mathbf{s}\triangleq[\mathbf{b}_{1}^{T}, \mathbf{b}_{2}^{T}, \ldots,
  \mathbf{b}_{K}^{T}]^{T} \in \mathbb{C}^{D \times 1}$ to the users,
where the transmit symbols are independent and identically distributed
(i.i.d.).  The symbol vector of user $k\in\mathcal{K}$,
$\mathbf{b}_{k}\triangleq [b_{1,k}, \ldots, b_{D_k, k}]^T \in
\mathbb{C}^{D_{k} \times 1}$, has zero-mean and covariance matrix
$\mathbb{E}[\mathbf{b}_{k} \mathbf{b}_{k}^{H}] = \mathbf{I}$, where
$D_k$ ($D_k \leq R_{r,k}$) denotes the number of data streams for user
$k$, and $D = \sum\limits_{k=1}^{K}D_{k}$ denotes the total number of
data streams. In this work, QPSK and $M$-ary square QAM symbol
constellations are adopted, although extensions to other types of
constellations are possible. For the QPSK case, the real and imaginary
parts of each entry of the signal vector $\mathbf{b}_k$ are uniformly
drawn from $\{\pm 1\}$. For the $M$-QAM modulation, each entry of
$\mathbf{b}_k$ is uniformly drawn from $\{F_m+jF_n: 1\leq m, n \leq
\sqrt{M}\}$, where integer $M$ is a perfect square and we define
$F_n=2n-\sqrt{M}-1$.  The transmitted symbol vector $\mathbf{s}$ is
first processed by a digital transmit beamforming matrix
$\mathbf{V}\triangleq [\mathbf{P}_1,\ldots, \mathbf{P}_K] \in
\mathbb{C}^{R_{t} \times D}$, where $\mathbf{P}_k\in\mathbb{C}^{R_t
  \times D_k}$ denotes the transmit beamforming matrix for user
$k$. The digital beamformer output is passed through the RF chains and
then processed by an analog transmit beamforming matrix $\mathbf{F}
\in \mathbb{C}^{N_{t} \times R_{t}}$ implemented by means of phase
shifters, i.e., $|[\bF]_{m, n}|=1$. The transmit power constraint at
the BS is given by \vspace{-1mm}
\begin{equation}       \label{st power}
	\mathbb{E} \{\| \mathbf{F} \mathbf{V} \mathbf{s} \|^2 \} = \| \mathbf{F} \mathbf{V} \|^2
 = P_T,  \vspace{-1mm}
\end{equation}
where  $P_T$ denotes the total transmit power budget.

The signal received at user $k$ is given by \vspace{-1mm}
\begin{equation}   \label{y}
	\mathbf{y}_k = \mathbf{H}_k \mathbf{F} \mathbf{V} \mathbf{s} + \mathbf{n}_k
               = \mathbf{H}_k \mathbf{F} \sum^{K}_{k=1}\mathbf{P}_{k} \mathbf{b}_k + \mathbf{n}_k, \vspace{-2mm}
\end{equation}
where $\mathbf{H}_{k} \in \mathbb{C}^{N_{r, k} \times N_{t}}$ denotes
the massive MIMO channel matrix between the BS and user $k$. As
discussed in \cite{2013Spatially} and \cite{2016Hybrid}, this matrix
is conveniently described by a geometry clustered channel model with
$N_{C}$ clusters and $N_{R}$ rays in each cluster due to the sparse
scattering property of massive MIMO channels. Considering a system with a
half-wave spaced uniform linear array at both the transmitter and the
receiver, we have without loss of generality \vspace{-1mm}
\begin{equation}
	\mathbf{H}_k = \sqrt{\dfrac{N_t N_{r, k}}{N_C N_R}} \sum_{i=1}^{N_C} \sum_{j=1}^{N_R} \alpha_{i,j} \mathbf{a}_r(\theta_{i,j}^r) \mathbf{a}_t(\theta_{i,j}^t)^H, \vspace{-1mm} \label{H}
\end{equation}
where $\alpha_{i,j}$ denotes the complex propagation gain of the
$j$-th ray in the $i$th cluster, while $\mathbf{a}_r(\theta_{i,j}^r)
\triangleq \dfrac{1}{\sqrt{N_{r, k}}}[1, e^{j\pi sin\theta_{i,j}^r},
  \ldots, e^{j\pi (N_r-1) sin\theta_{i,j}^r}]^T$ and
$\mathbf{a}_t(\theta_{i,j}^t) \triangleq \dfrac{1}{\sqrt{N_t}}[1,
  e^{j\pi sin\theta_{i,j}^t}, \ldots, e^{j\pi (N_t-1)
    sin\theta_{i,j}^t}]^T$ denote the corresponding normalized
response vectors of the transmit and receive antenna arrays, with
$\theta^r_{i,j}$ and $\theta^t_{i,j}$ denoting the angles of arrival
and departure, respectively. The term $\mathbf{n}_k \in
\mathbb{C}^{N_{r, k} \times 1}$ in \eqref{y} represents the additive
noise at user $k$, modeled as a complex circularly symmetric Gaussian
vector with zero-mean and covariance matrix $\mathbb{E}[\mathbf{n}_k
  \mathbf{n}_k^H] = \sigma_k^2 \mathbf{I}$.

Similar to the BS, the hybrid AD receiver at user $k$ is comprised of
an analog receive beamforming matrix $\mathbf{U}_{k} \in
\mathbb{C}^{R_{r,k} \times N_{r}}$, followed by a digital receive
beamforming matrix $\mathbf{W}_{k}\triangleq [\mathbf{w}_{1, k},
  \ldots, \mathbf{w}_{D_{k}, k}] \in \mathbb{C}^{R_{r,k} \times D_k}$,
where $\mathbf{w}_{i, k} \in \mathbb{C}^{R_{r,k} \times 1}$ and $i\in
\mathcal{D}_k\triangleq\{1, \ldots, D_k\}$.
Accordingly, the output  signal vector $\tilde{\mathbf{b}}_k \triangleq [\tilde{b}_{1, k}, \ldots, \tilde{b}_{D_k, k}]^T\in \mathbb{C}^{D_k\times 1}$
of the hybrid receiver at user $k$ can be expressed as \vspace{-1mm}
\begin{equation}
 \tilde{\mathbf{b}}_k = \mathbf{W}^H_k \mathbf{U}_k \mathbf{y}_k= \mathbf{W}^H_k \mathbf{U}_k (\mathbf{H}_k \mathbf{F} \sum^{K}_{k=1}\mathbf{P}_{k} \mathbf{b}_k + \mathbf{n}_k), \vspace{-1mm}
\end{equation}
\noindent
where the entries of the analog beamforming matrices obey the unit-modulus constraints, i.e.,
$|[\mathbf{U}_k]_{m, n}| = 1$,  $\forall k\in\mathcal{K}$.

Finally, the estimate of the transmitted symbol vector $\mathbf{b}_k$ at the output of receiver $k$ is \vspace{-1mm}
	\begin{equation}   \label{b_hat}
	\hat{\mathbf{b}}_k  = \mathcal{Q}\{\tilde{\mathbf{b}}_k\},\vspace{-1mm}
	\end{equation}
where $\mathcal{Q}\{\cdot\}$ is the quantization operation for the given modulation.

The basic problem in implementing the hybrid AD transceiver scheme is
how to effectively design the beamforming matrices $\{\mathbf{P}_k,
\mathbf{W}_k, \mathbf{U}_k, \mathbf{F} \}$ to detect the transmitted
symbols accurately under the total transmit power constraint at the BS
and the unit-modulus constraint imposed on each element of the analog
RF beamforming matrices at the BS and user sides.

\section{Proposed MSER-Based GD Algorithm for Hybrid AD Transceiver Design}
\label{Section3:gradient-descent}

In this section, we focus on the QPSK case to formulate the MSER
criterion mathematically.  The resultant problem is very difficult to
tackle due to the highly nonlinear objective function and constraints.
We then develop a MSER-based GD iterative algorithm for finding the
stationary points, which will serve as the basis in the elaboration of
our proposed deep unfolding NN.

\vspace{-2mm}
\subsection{MSER Criterion}

The decisions for the detection of the real and imaginary parts of
element $i\in\mathcal{D}_k$ in the $k$-th user's symbol vector
$\mathbf{b}_k=[b_{1,k},...,b_{D_k,k}]^T$ are made as \vspace{-1mm}
\begin{equation}
\begin{split}
	\Re\{\hat{b}_{i, k} \} = \left\{
\begin{aligned}
		& +1, \qquad \textrm{if} \ \Re\{\tilde{b}_{i, k} \} \geq 0 \\
		& -1, \qquad \textrm{if} \ \Re\{\tilde{b}_{i, k} \} < 0,
\end{aligned}
\right.
	\\
	\Im\{\hat{b}_{i, k} \} =  \left\{
\begin{aligned}
		& +1, \qquad \textrm{if} \ \Im\{\tilde{b}_{i, k} \} \geq 0 \\
		& -1, \qquad \textrm{if} \ \Im\{\tilde{b}_{i, k} \} < 0.  \vspace{-1mm}
\end{aligned}
\right.
\end{split}
\end{equation}

For a given desired symbol $b_{i, k}$, there exists $N_b = 4^{D-1}$
legitimate combinations of the multi-user interference symbols $\{
b_{i, k'}, i\in \mathcal{D}_{k^{'}}, k' \in \mathcal{K}, k' \neq k \}$
and self-interference symbols $\{b_{i^{'}, k}, i' \in \mathcal{D}_k,
i' \neq i \}$. We define the set of all the possible transmitted
symbol vectors as \vspace{-1mm}
\begin{equation}  \label{X}
	\mathcal{X} \triangleq \{\mathbf{s}^1, \mathbf{s}^2, \ldots, \mathbf{s}^{N_b} \},
\end{equation}
where $\mathbf{s}^q = [(\mathbf{b}_{1}^{q})^T, (\mathbf{b}_{2}^{q})^T,
  \ldots, (\mathbf{b}_{K}^{q})^T]^T$, $q \in \mathcal{N}_b\triangleq
\{1, \ldots, N_b\}$, and we assume an equiprobable model for the $N_b$
possible transmit vectors $\mathbf{s}^q$. The noise-free component for
the $i$-th element of the hybrid receiver's output at user $k$ is from
the set \vspace{-1mm}
\begin{equation}
	\mathcal{Y}_{i,k} \triangleq \{\bar{b}^q_{i, k} = \mathbf{w}^H_{i, k} \mathbf{U}_k \mathbf{H}_k \mathbf{F} \mathbf{V} \mathbf{s}^q, \, q\in \mathcal{N}_b \}. \vspace{-1mm}
\end{equation}
Due to the Gaussian distribution for the additive noise at the
receiver, and invoking the law of total probability, we can express
the probability density function (PDF) for the real part of the hybrid
receiver's output, given the symbol $b_{i, k}$, as \vspace{-1mm}
\begin{equation}  \label{pdf_old}
	\begin{split}
		\!\!\!\!\!f(x|b_{i, k}) =   &\frac{1}{ N_b  \sqrt{2\pi \mathbf{w}_{i, k}^H \mathbf{U}_k \mathbf{H}_k \mathbf{F} \mathbf{V} \mathbf{V}^H \mathbf{F}^H \mathbf{H}_k^H  \mathbf{U}_k^H \mathbf{w}_{i, k}} \sigma_n \!}
		\\
		&\sum_{q=1}^{N_b}  e^{ -\frac{\left|x- \Re \{\bar{b}^q_{i, k}\} \right|^2}{ 2\mathbf{w}^H_{i, k} \mathbf{U}_k \mathbf{H}_k \mathbf{F} \mathbf{V} \mathbf{V}^H \mathbf{F}^H \mathbf{H}_k^H \mathbf{U}_k^H \mathbf{w}_{i, k} \sigma_n^2 } },
	\end{split}
\end{equation}
where $\bar{b}^q_{i, k} \in \mathcal{Y}$.  Due to the large value of
$N_b$, it will not be possible to consider the sum over all the
possible transmitted symbol vectors.  In practice, the PDF of the
receiver's output should be approximated based on a block of
experimental samples. Specifically, with the aid of kernel density
estimation \cite{MSERQPSK}, we randomly select $J$ different transmit
symbol vectors from the set $\mathcal{X}$ in \eqref{X}, indexed with
$q_j$ for $j\in\{1,...,J\}$, and employ a constant kernel width
$\varrho$ to replace the term $\sqrt{\mathbf{w}^H_{i, k} \mathbf{U}_k
  \mathbf{H}_k \mathbf{F} \mathbf{V} \mathbf{V}^H \mathbf{F}^H
  \mathbf{H}_k^H \mathbf{U}_k^H \mathbf{w}_{i, k}}\sigma_n$ appearing
in \eqref{pdf_old} for complexity reduction. The parameter $\varrho$
is related to the noise standard deviation and is selected based on
separate simulations \cite{MSERQPSK}.  The block-data kernel estimate
of the true PDF for the real part of the receiver output
is \vspace{-1mm}
\begin{equation}   \label{pdf}
f(x|b_{i, k}) =  \frac{1}{J \sqrt{2\pi} \varrho}\,  \sum_{{j}=1}^{J} e^{ -\frac{|x- \Re \{\bar{b}^{q_j}_{i, k}\} |^2}{2 \varrho^2} }. \vspace{-0mm}
\end{equation}
The PDF for the imaginary part of the receiver's output can be
obtained in the same way.

Furthermore, the  SER of $b_{i, k}$ is given by \vspace{-1mm}
\begin{equation}    \label{initial SER}
	\mathcal{P}_e(b_{i, k}) = \mathcal{P}_e^R(b_{i, k}) + \mathcal{P}_e^I(b_{i, k}) - \mathcal{P}_e^R(b_{i, k}) \mathcal{P}_e^I(b_{i, k}), \vspace{-0mm}
\end{equation}
where $\mathcal{P}_e^R(b_{i, k}) \triangleq \mathrm{Prob}\left\{\Re\{\hat{b}_{i, k}\} \neq \Re\{b_{i, k} \} \right\}$ and $\mathcal{P}_e^I(b_{i, k}) \triangleq \mathrm{Prob}\left\{\Im\{\hat{b}_{i, k}\} \neq \Im\{b_{i, k} \} \right\}$ represent the real-part and imaginary-part SER, respectively.
Based on \eqref{pdf}, we have \vspace{-1mm}
\begin{equation}     \label{SER_R}
	\mathcal{P}_e^R(b_{i, k}) =  \frac{1}{J \sqrt{\pi}}  \sum \limits_{{j}=1}^J \int_{- \infty}^{-\frac{\Re \{\bar{b}^{q_j}_{i, k}\} \Re\{b_{i, k}\}}{\sqrt{2}\varrho}} \ e^{-s^2} ds,
\end{equation} \vspace{-3mm}
\begin{equation}      \label{SER_I}
	\mathcal{P}_e^I(b_{i, k}) =  \frac{1}{J \sqrt{\pi}}  \sum \limits_{{j}=1}^J \int_{- \infty}^{-\frac{\Im \{\bar{b}^{q_j}_{i, k}\} \Im\{b_{i, k}\}}{\sqrt{2}\varrho}} \ e^{-s^2} ds.\vspace{-1mm}
\end{equation}
According to the discussion in \cite{MSERQAM}, we can drop the product term $\mathcal{P}_e^R(b_{i, k}) \mathcal{P}_e^I(b_{i, k})$ and focus on  the  upper bound $\tilde{\mathcal{P}}_e(b_{i, k})\triangleq \mathcal{P}_e^R(b_{i, k}) + \mathcal{P}_e^I(b_{i, k})$  for reducing the computational complexity. Indeed, for small values of SER, $\tilde{\mathcal{P}}_e(b_{i, k})$ is very close to the true SER $\mathcal{P}_e(b_{i, k})$, i.e., the bound is tight.
We aim to minimize the upper bound of the overall SER over all user data by jointly optimizing the beamforming matrices $\{\mathbf{P}_k, \mathbf{W}_{k}, \mathbf{U}_k, \mathbf{F} \}$, where index $k$ runs through $\mathcal{K}$. 
 Accordingly, the problem is formulated as \vspace{-1mm}
\begin{subequations}   \label{problem_formulation}
	\begin{align}
		\!\!\!\!\!\!\!\!\!\min \limits_{\{\mathbf{P}_k, \mathbf{W}_{k}, \atop \mathbf{U}_k, \mathbf{F}\}}    &\sum^{K}_{k=1}\sum^{D_k}_{i=1} \tilde{\mathcal{P}}_e(b_{i, k}) \triangleq \sum^{K}_{k=1}\sum^{D_k}_{i=1}(\mathcal{P}_e^R(b_{i, k}) + \mathcal{P}_e^I(b_{i, k})) \vspace{-5mm} \label{min} \\
		\st \ &\| \mathbf{F} \mathbf{V} \|^2 = P_T,    \label{st_power}\\
		\ &|[\mathbf{F}]_{m, n}| = 1, \,
		|[\mathbf{U}_k]_{m, n}| = 1, \quad \, \forall k, m, n,  \label{st_unit2}
	\end{align}
\end{subequations}
where   \eqref{st_power}  and  \eqref{st_unit2} are  the transmit power  and  constant modulus constraints, respectively.

\vspace{-4mm}
\subsection{MSER-based GD Joint Beamforming Design Algorithm}

In the following, we introduce the proposed MSER-based GD joint
beamforming design algorithm.  To tackle the unit-modulus constraints
in \eqref{st_unit2}, we define the analog beamforming phase matrices
$\bm{\theta}_{U_k} \triangleq \angle \mathbf{U}_k$ and $\bm{\theta}_F
\triangleq \angle \mathbf{F}$.  Consequently, $\mathbf{U}_k$ and
$\mathbf{F}$ can be obtained by $\mathbf{U}_k = \exp(j
\bm{\theta}_{U_k})$ and $\mathbf{F} = \exp(j \bm{\theta}_F)$,
respectively, where the function $\exp(\cdot)$ is applied element-wise.
Furthermore, to guarantee the transmit power
constraint, we shall scale the overall digital transmit beamforming
matrix $\mathbf{V} = [\mathbf{P}_1, \ldots, \mathbf{P}_K]$ at the end
of each iteration of the proposed MSER-based GD algorithm.  Therefore,
let us consider the following unconstrained problem for
simplicity: \vspace{-1.5mm}
\begin{equation}
	\min \limits_{\{\mathbf{P}_k, \mathbf{W}_{k}, \bm{\theta}_{U_k}, \bm{\theta}_F\}} \, \sum^{K}_{k=1}\sum^{D_k}_{i=1} \tilde{\mathcal{P}}_e(b_{i, k})\triangleq \sum^{K}_{k=1}\sum^{D_k}_{i=1}(\mathcal{P}_e^R(b_{i, k}) + \mathcal{P}_e^I(b_{i, k})). \vspace{-1mm}  \label{eq:mserobjective}
\end{equation}

Based on the objective function of MSER, the gradient of
$\tilde{\mathcal{P}}_e(b_{i, k})$ with respect to (w.r.t.) the hybrid
beamforming matrices is given by $\nabla\tilde{\mathcal{P}}_e(b_{i,
  k}) \triangleq \nabla \mathcal{P}_e^R(b_{i, k}) + \nabla
\mathcal{P}_e^I(b_{i, k})$, where $\nabla \mathcal{P}_e^R(b_{i, k})$
and $\nabla \mathcal{P}_e^I(b_{i, k})$ denote the gradients w.r.t. the
real part and the imaginary part, respectively.  Let us first focus on
deriving the gradients w.r.t. the real parts of $\{\mathbf{P}_k,
\mathbf{W}_{k}, \mathbf{U}_k, \mathbf{F} \}$.  Specifically, computing
the gradient of \eqref{SER_R}, as in \cite{matrix_analysis}, we
obtain, \vspace{-3mm}
	\begin{equation}     \label{delta_P}
	\begin{split}
		\nabla_{\mathbf{P}^{*}_k} \mathcal{P}_e^R  =  &-\frac{1} {J \sqrt{2\pi} \varrho} \sum \limits_{{j}=1}^J \, e^{-\frac{|\Re \{\bar{b}^{q_j}_{i,k}\}|^2} {2\varrho^2}} \ \Re \{ b_{i,k} \} \\ &\mathbf{F}^H \mathbf{H}_k^H \mathbf{U}_k^H \mathbf{W}_k (\mathbf{b}_k^{q_j})^H,
	\end{split}
	\end{equation}\vspace{-2mm}
	\begin{equation}    \label{delta_W}
	\begin{split}
		\nabla_{\mathbf{w}^{*}_{i,k}} \mathcal{P}_e^R  = &-\frac{1} {J \sqrt{2\pi} \varrho} \sum \limits_{{j}=1}^J \, e^{-\frac{|\Re \{\bar{b}^{q_j}_{i,k}\}|^2} {2\varrho^2}} \ \Re \{ b_{i,k} \} \\ &\mathbf{U}_k \mathbf{H}_k \mathbf{F} (\mathbf{V} \mathbf{s}^{q_j}),
	\end{split}
	\end{equation}\vspace{-2mm}
	\begin{equation}     \label{delta_U}
	\begin{split}
		\nabla_{\mathbf{U}^{*}_k} \mathcal{P}_e^R  = &-\frac{1} {J \sqrt{2\pi} \varrho} \sum \limits_{{j}=1}^J \, e^{-\frac{|\Re \{\bar{b}^{q_j}_{i,k}\}|^2} {2\varrho^2}} \ \Re \{ b_{i,k} \} \\ & \mathbf{w}_{i,k} (\mathbf{V} \mathbf{s}^{q_j})^H  \mathbf{F}^H \mathbf{H}_k^H,
	\end{split}	
	\end{equation}\vspace{-2mm}
	\begin{equation}     \label{delta_F}
	\begin{split}
		\nabla_{\mathbf{F}^{*}} \mathcal{P}_e^R   =  &-\frac{1} {J \sqrt{2\pi} \varrho} \sum \limits_{{j}=1}^J \, e^{-\frac{|\Re \{\bar{b}^{q_j}_{i,k}\}|^2} {2\varrho^2}} \ \Re \{ b_{i,k} \} \\ & \mathbf{H}_k^H \mathbf{U}_k^H \mathbf{w}_{i,k}   \,  (\mathbf{V} \mathbf{s}^{q_j})^H,
	\end{split}	
	\end{equation}
and we define $\nabla_{\mathbf{W}^{*}_{k}} \mathcal{P}_e^R \triangleq [\nabla_{\mathbf{w}^{*}_{1,k}} \mathcal{P}_e^R, \ldots,  \nabla_{\mathbf{w}^{*}_{D_{k},k}}\mathcal{P}^R_e]$.

The gradients w.r.t. the analog beamforming phase matrices $\bm{\theta}_{U_k}$ and $\bm{\theta}_F $ can be obtained as \vspace{-1mm}
\begin{equation}
	\begin{split}
	\nabla_{\bm{\theta}_{U_k}} \mathcal{P}^{R}_e & = \Re\{\nabla_{\mathbf{U}_k} \mathcal{\tilde{P}}_e \circ j \mathbf{U}_k - \nabla_{\mathbf{U}_k^*} \mathcal{\tilde{P}}_e \circ j \mathbf{U}_k^*\}
	\\
	& = -\nabla_{\mathbf{U}_k} \mathcal{P}^{R}_e \! \circ \mathbf{U}_k^I - \nabla_{\mathbf{U}_k} \mathcal{P}^{I}_e \! \circ \mathbf{U}_k^R
	\\ & \quad \ + \nabla_{\mathbf{U}_k^{*}} \mathcal{P}^{R}_e \circ (\mathbf{U}_k^{I})^{*} + \nabla_{\mathbf{U}_k^*} \mathcal{P}^{I}_e \circ (\mathbf{U}_k^{R})^{*},          \label{delta_theu}
	\end{split}
\end{equation} \vspace{-1mm}
\begin{equation}
	\begin{split}
		\nabla_{\bm{\theta}_F} \mathcal{P}^{R}_e & = \Re\{\nabla_{\mathbf{F}} \mathcal{\tilde{P}}_e \circ j \mathbf{F} - \nabla_{\mathbf{F}^*} \mathcal{\tilde{P}}_e \circ j \mathbf{F}^*\}
		\\
		& = -\nabla_{\mathbf{F}} \mathcal{P}^{R}_e \circ \mathbf{F}^I - \nabla_{\mathbf{F}} \mathcal{P}^{I}_e \circ \mathbf{F}^R
		\\ & \quad \ + \nabla_{\mathbf{F}^*} \mathcal{P}^{R}_e \circ (\mathbf{F}^{I})^{*} + \nabla_{\mathbf{F}^*} \mathcal{P}^{I}_e \circ (\mathbf{F}^{R})^{*}.           \label{delta_thef}
	\end{split}
\end{equation}
The gradients w.r.t. the imaginary parts of the beamforming matrices
can be calculated similarly, thus the details are omitted for brevity.

\begin{figure}[t]
	\centering
	\includegraphics[width=0.5\textwidth]{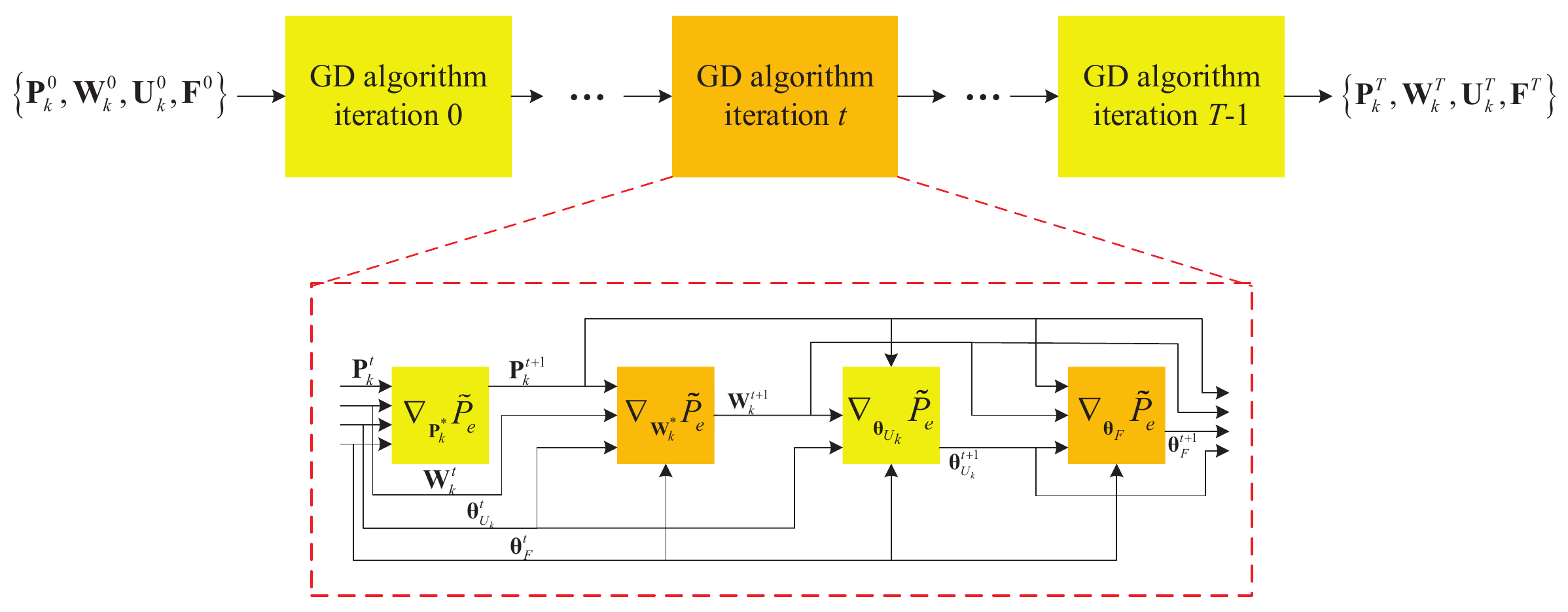}
	\caption{Structure of the proposed MSER-based iterative GD algorithm.}
	\label{GD structure}
\end{figure}

The hybrid beamforming matrices are jointly optimized based on the
MSER criterion. The GD update equations are obtained by substituting
the gradients \eqref{delta_P}--\eqref{delta_thef} in the following
expressions \vspace{-1mm}
\begin{equation}
		\mathbf{P}_k^{t+1} = \mathbf{P}_k^{t} - \mu_P \nabla_{\mathbf{P}^{*}_k} \tilde{\mathcal{P}}_e, \vspace{-1mm}     \label{update_p}
	\end{equation}
	\begin{equation}
		\ \ \ \mathbf{W}_k^{t+1} = \mathbf{W}_k^{t} - \mu_W \nabla_{\mathbf{W}^{*}_k} \tilde{\mathcal{P}}_e,   \vspace{-1mm}  \label{update_w}
	\end{equation}
	\begin{equation}
		\quad \bm{\theta}_{U_k}^{t+1} = \bm{\theta}_{U_k}^{t} - \mu_{\theta_U} \nabla_{\bm{\theta}_{U_k}} \tilde{\mathcal{P}}_e,  \vspace{-1mm}       \label{update_U}
	\end{equation}
	\begin{equation}
		\ \bm{\theta}_F^{t+1} = \bm{\theta}_F^{t} - \mu_{\theta_F} \nabla_{\bm{\theta}_F} \tilde{\mathcal{P}}_e,    \vspace{-1mm} \label{update_F}
	\end{equation}
where $\{\mu_P, \mu_W, \mu_{\theta_U}, \mu_{\theta_F}\}$ denote the step sizes employed in the GD iterations, and $t\in\{0, 1, \ldots\}$  is the iteration index.
The beamforming matrices are updated alternately until  a certain convergence criterion is met.
The iterative structure of the MSER-based GD algorithm structure is shown in Fig. \ref{GD structure}.

To guarantee the transmit power constraint \eqref{st_power}, at the
end of each GD iteration, the overall digital transmit beamforming
matrix $\mathbf{V}$ needs to be scaled as follows: \vspace{-1mm}
	\begin{equation}     \label{scaling}
		\mathbf{V} \leftarrow \frac{\sqrt{P_T}}{\| \mathbf{F}\mathbf{V} \|} \, \mathbf{V}.
	\end{equation}
The details of the proposed MSER-based GD hybrid beamforming design
algorithm are summarized in Algorithm \ref{tab:GD algorithm}.  The
latter is devised to start its operation in the training mode, where a
known training transmit symbol sequence is employed, and then switch
to the decision-directed mode, wherein the estimated symbols are used
for computation.

\begin{algorithm}[t]
	\caption{Gradient-descent algorithm for joint hybrid  beamforming design}
	\label{tab:GD algorithm}
	\begin{small}
		\begin{spacing}{0.9}
			\begin{algorithmic}[1]  \vspace{1mm}
				\State \parbox[t]{\dimexpr\linewidth-\algorithmicindent-\algorithmicindent}{Set the tolerance of accuracy $\epsilon$, the maximum number of iterations $T$, the size of the random sample $J$, and the step sizes $\{\mu_P, \mu_W, \mu_{\theta_U}, \mu_{\theta_F}\}$. Set the iteration index to $t=0$.}  \vspace{1.5mm}
				\State \parbox[t]{\dimexpr\linewidth-\algorithmicindent-\algorithmicindent}{Initialize $\mathbf{P}_k$ to satisfy the power constraint. Initialize $\{\mathbf{W}_k, \bm{\theta}_{U_k}, \bm{\theta}_F \}$.}    \vspace{1mm}
				\Repeat  \vspace{1mm}
				\State \parbox[t]{\dimexpr\linewidth-\algorithmicindent-\algorithmicindent}{ Update $\mathbf{P}_k^{t+1}$ with fixed $\{\mathbf{W}_k^{t}, \bm{\theta}_{U_k}^{t}, \bm{\theta}_F^{t} \}$, $\forall k\in\mathcal{K}$, according to \eqref{update_p}.
				}  \vspace{1mm}
				\State \parbox[t]{\dimexpr\linewidth-\algorithmicindent-\algorithmicindent}{Update  $\mathbf{W}_k^{t+1}$ with fixed $\{\mathbf{P}_k^{t+1}, \bm{\theta}_{U_k}^{t}, \bm{\theta}_F^{t} \}$, $\forall k\in\mathcal{K}$, according to \eqref{update_w}.}  \vspace{1mm}
				\State \parbox[t]{\dimexpr\linewidth-\algorithmicindent-\algorithmicindent}{ Update $\bm{\theta}_{U_k}^{t+1}$ with fixed $\{\mathbf{P}_k^{t+1}, \mathbf{W}_k^{t+1}, \bm{\theta}_F^{t} \}$, $\forall k\in\mathcal{K}$, according to \eqref{update_U}.
				}  \vspace{1mm}
				\State \parbox[t]{\dimexpr\linewidth-\algorithmicindent-\algorithmicindent}{ Update $\bm{\theta}_F^{t+1}$ with fixed $\{\mathbf{P}_k^{t+1}, \mathbf{W}_k^{t+1}, \bm{\theta}_{U_k}^{t+1}  \}$, according to \eqref{update_F}.
				}  \vspace{1mm}
				\State \parbox[t]{\dimexpr\linewidth-\algorithmicindent-\algorithmicindent}{Scale $\mathbf{P}_k^{t+1}$ based on \eqref{scaling} to meet the transmit power constraint.} \vspace{1mm}
				\State Update the iteration index: $t=t+1$.  \vspace{1mm}
				\Until{The objective function meets chosen convergence criterion or $t > {T}$.}   \vspace{0.5mm}
			\end{algorithmic}
		\end{spacing}
	\end{small}
\end{algorithm}

\vspace{-2mm}
\section{ Proposed Deep-Unfolding NN for Hybrid AD Transceiver Design}
\label{Section4:deep-unfolding}

The conventional MSER-based GD algorithms usually provide very slow
convergence speed and therefore require a large number of
iterations. Moreover, they suffer from performance degradation in the
presence of channel state information (CSI) errors.  To address these
issues, we propose a deep-unfolding NN to jointly design the hybrid AD
transceiver, where a small number of NN layers with trainable
parameters can be employed while maintaining satisfactory
performance. The proposed NN structure is inspired by the MSER-based
GD algorithm, where the iterative algorithm is unfolded into a
multi-layer structure and a number of adjustable step size and bias
parameters are introduced in the FP.  In the training stage, the
relationship between the gradients of adjacent layers is derived
according to the GCR in the BP. We then calculate the gradients
w.r.t. the trainable parameters layer by layer and update these
parameters based on the stochastic gradient descent (SGD)
algorithm. In the testing stage, we perform the FP process based on
the trained parameters for computing the AD beamforming matrices. The
details of the FP and BP in the proposed NN are presented as follows.

\vspace{-3mm}
\subsection{Forward Propagation}
\vspace{-0mm}

In this subsection, we describe the structure of the proposed
deep-unfolding NN which is induced by the MSER-based GD algorithm
developed in Section \ref{Section3:gradient-descent}. In the iteration
of the latter algorithm, the step sizes used for updating the hybrid
AD beamforming matrices, i.e., $\{\mu_P, \mu_W, \mu_{\theta_U},
\mu_{\theta_F}\}$ greatly affect the SER performance; furthermore, are
usually determined based on experience and experiments. Therefore, in
layer $l\in \{0, \ldots, L-1\}$ of the proposed NN, where $L$ denotes
the number of layers, we introduce the trainable matrix parameters
$\{\bm{\alpha}_{P_k}^l, \bm{\alpha}_{W_k}^l,
\bm{\alpha}_{\theta_{U_k}}^l, \bm{\alpha}_{\theta_F}^l \}$ as the
learning rates to replace the step sizes of the MSER-based GD
algorithm. Recall that the term $\sqrt{\mathbf{w}^H_{i,k} \mathbf{U}_k
  \mathbf{H}_k \mathbf{F} \mathbf{V} \mathbf{V}^H \mathbf{F}^H
  \mathbf{H}_k^H \mathbf{U}_k^H \mathbf{w}_{i,k}}\sigma_n$ in
\eqref{pdf_old} is set as a constant kernel width $\varrho$ in the
MSER-based GD algorithm, which may cause performance loss. Hence, we
replace $\varrho$ in \eqref{delta_P}--\eqref{delta_F} by the set of
trainable parameters $\{\rho_{P_k}^l, \rho_{W_k}^l, \rho_{U_k}^l,
\rho_F^l \}$. Moreover, to increase the degrees of freedom for the
parameters, we introduce the trainable offset matrix parameters
$\{\mathbf{O}_{P_k}^l, \mathbf{O}_{W_k}^l,
\mathbf{O}_{\theta_{U_k}}^l, \mathbf{O}_{\theta_F}^l \}$ for the
computation of the beamforming matrices. The update expressions for
the proposed deep-unfolding NN are given by
\vspace{-2mm}
	\begin{equation}
		\mathbf{P}_k^{l+1} = \mathbf{P}_k^{l} - \bm{\alpha}_{P_k}^l \circ \nabla_{\mathbf{P}_k^{*}} \tilde{\mathcal{P}}_e^{l} + \mathbf{O}_{P_k}^l,  \vspace{-1mm}  \label{update_p_new}
	\end{equation}
	\begin{equation}
		\quad \ \mathbf{W}_k^{l+1} = \mathbf{W}_k^{l} - \bm{\alpha}_{W_k}^l \circ \nabla_{\mathbf{W}_k^{*}} \tilde{\mathcal{P}}_e^{l} + \mathbf{O}_{W_k}^l,     \vspace{-1mm}  \label{update_w_new}
	\end{equation}
	\begin{equation}
		\quad \ \bm{\theta}_{U_k}^{l+1} = \bm{\theta}_{U_k}^{l} - \bm{\alpha}_{\theta_{U_k}}^l \circ \nabla_{\bm{\theta}_{U_k} } \tilde{\mathcal{P}}_e^{l} + \mathbf{O}_{\theta_{U_k}}^l,   \vspace{-1mm}     \label{update_U_new}
	\end{equation}
	\begin{equation}
		\bm{\theta}_F^{l+1} = \bm{\theta}_F^{l} - \bm{\alpha}_{\theta_F}^l \circ \nabla_{\bm{\theta}_F } \tilde{\mathcal{P}}_e^{l} + \mathbf{O}_{\theta_F}^l,   \vspace{-0mm}  \label{update_F_new}
	\end{equation}
where  $\{\bm{\alpha}_{P_k}^l, \mathbf{O}_{P_k}^l\}\in\mathbb{C}^{R_t \times D_k}$, $\{\bm{\alpha}_{W_k}^l, \mathbf{O}_{W_k}^l\}\in\mathbb{C}^{R_{r,k} \times D_k}$, $\{\bm{\alpha}_{\theta_{U_k}}^l, \mathbf{O}_{\theta_{U_k}}^l\}\in\mathbb{C}^{R_{r,k} \times N_{r,k}}$, $\{\bm{\alpha}_{\theta_F}^l, \mathbf{O}_{\theta_F}^l\}\in\mathbb{C}^{N_t \times R_t}$, and $\{\nabla_{\mathbf{P}_k^{*}} \tilde{\mathcal{P}}_e^{l}, \nabla_{\mathbf{W}_k^{*}} \tilde{\mathcal{P}}_e^{l}, \nabla_{\bm{\theta}_{U_k} } \tilde{\mathcal{P}}_e ^{l}, \nabla_{\bm{\theta}_F } \tilde{\mathcal{P}}_e ^{l}\}$ denote the gradients w.r.t. the beamforming matrices in the $l$-th layer.

The structure of the proposed deep-unfolding NN induced by the
MSER-based GD algorithm is illustrated in Fig. \ref{deep-unfolding
  framework}. Compared with Fig. \ref{GD structure}, we can see that
this structure is developed by unfolding the iterative GD algorithm
into a multi-layer, comprised of $L$ successive layers (top). The
enlarged part in the red solid rectangle (middle) presents the common
details of each layer in the deep-unfolding NN, where the operations
$\mathcal{P}(\vartheta)$, $\mathcal{W}(\vartheta)$,
$\mathcal{U}(\vartheta)$, and $\mathcal{F}(\vartheta)$ represent
\eqref{update_p_new}--\eqref{update_F_new},
respectively. Specifically, in layer $l \in \{0, \ldots, L-1\}$, we
first update the digital beamforming matrices $\mathbf{P}_k$ and
$\mathbf{W}_k$ successively, followed by the analog beamforming phase
matrices $\bm{\theta}_{U_k}$ and $\bm{\theta}_F$. For each one of
these updates, the enlarged diagram in the red dotted rectanlge
(bottom) illustrates how the GD updates make use of the additional
parameters $\bm{\alpha}_X^l$, $\rho_X^l$, and $\mathbf{O}_X^l$, where
symbol $X \in \{P_k, W_k, \theta_{U_k}, \theta_F\}$. The beamforming
matrices in the last layer are served as the NN outputs and are
conveyed to the loss function. Since $\bm{\theta}_F$ is the last
updated parameter matrix, the iteration rule of $\bm{\theta}_F^L$
adopts the original formula \eqref{update_F} without introducing
additional parameters.
\begin{figure}[t]
	\centering
	\includegraphics[width=0.5\textwidth]{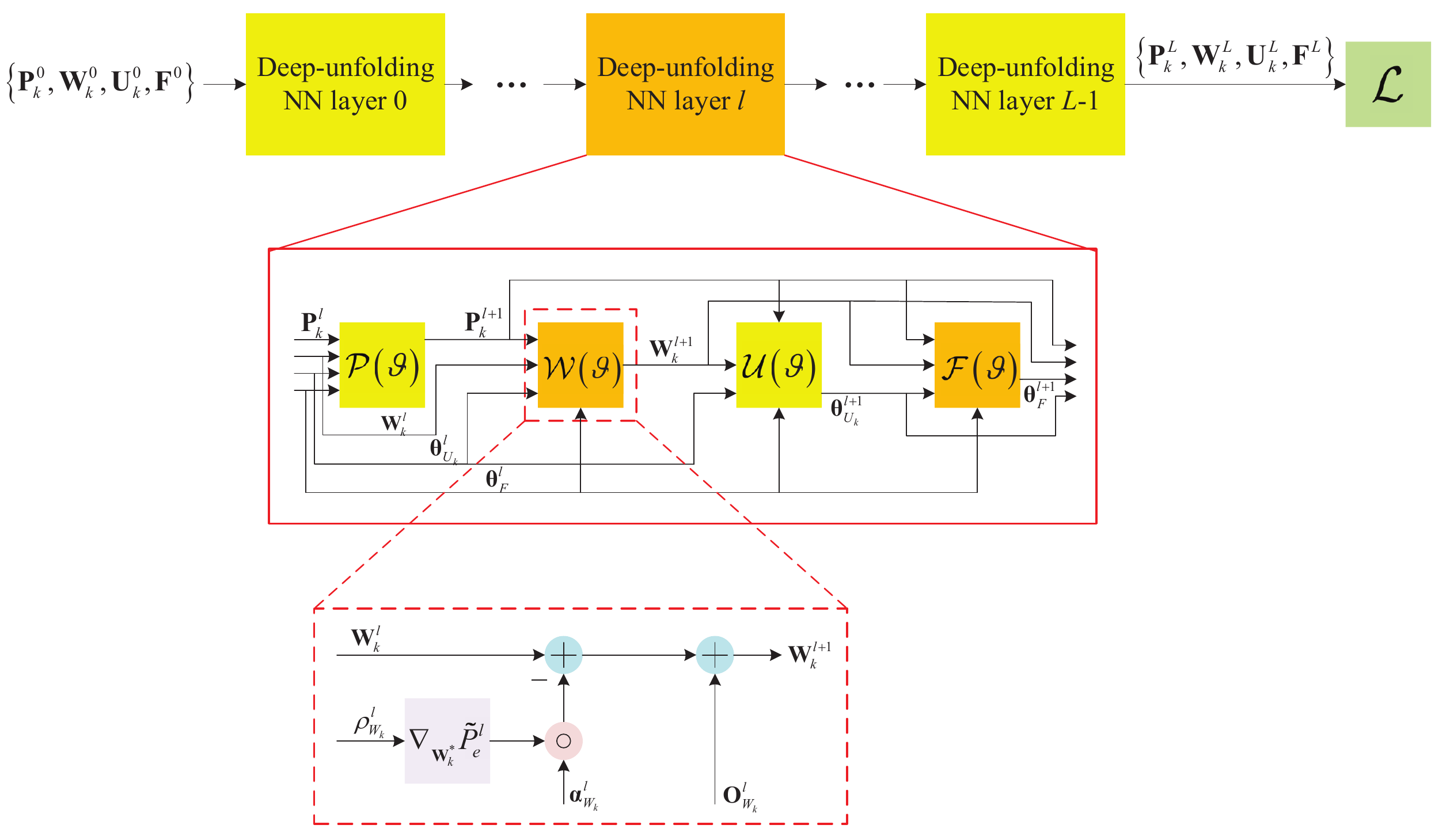}
	\caption{Structure of the proposed deep-unfolding NN induced by the MSER-based GD algorithm.}
	\label{deep-unfolding framework}
\end{figure}

Since the channel matrices $\mathbf{H}_k$ are random by nature, the
final loss function incorporates an expectation operation
$\mathbb{E}_\mathbf{H}$ over the ensemble of channel matrices.  Hence,
the loss function $\mathcal{L}$ in the top-right corner of
Fig. \ref{deep-unfolding framework} is modified as
\begin{equation}        \label{objective function}
\begin{split}
	\mathcal{L} = \sum \limits_{k=1}^K \mathbb{E}_{\mathbf{H}} &\Bigg\{ \frac{1}{J \sqrt{\pi}} \sum \limits_{i=1}^{D_k} \sum \limits_{{j}=1}^J \Big(\int_{- \infty}^{-\frac{\Re \{\bar{b}^{q_j}_{i,k}\} \Re\{b_{i,k}\}}{\sqrt{2}\rho}} e^{-s^2} \! \ ds
	\\
	&+ \int_{- \infty}^{-\frac{\Im \{\bar{b}^{q_j}_{i,k}\} \Im\{b_{i,k}\}}{\sqrt{2}\rho}} e^{-s^2} \! \ ds \Big) \Bigg\}.
\end{split}	
\end{equation}
As mentioned in Section \ref{Section3:gradient-descent}, $\mathbf{V} =
[\mathbf{P}_1, \ldots, \mathbf{P}_K]$ is scaled based on
\eqref{scaling} in each layer to satisfy the transmit power
constraint, which also helps avoid gradient explosion.

\vspace{-4mm}
\subsection{Back Propagation}
\label{deep-unfolding-back}

Since the conventional platforms for implementation and training of NN
(e.g. Pytorch or Tensorflow) are not designed to handle loss functions
in the form of integrals, we seek to propose a novel method to compute
the gradients in closed-form, which is more accurate and efficient. In
the BP, we derive the recursive relation between the gradients of
adjacent layers based on the GCR which is given in Appendix
\ref{appendixA}, and then compute the gradients w.r.t. the trainable
parameters.

Let $\{\mathbf{G}_{P_k}^l, \mathbf{G}_{W_k}^l, \mathbf{G}_{U_k}^l,
\mathbf{G}_{U^*_k}^l, \mathbf{G}_F^l, \mathbf{G}_{F^*}^l \}$ denote
the gradients w.r.t. the hybrid AD beamforming matrices in the $l$-th
layer. By taking the derivative of \eqref{objective function}, we can
calculate the gradients w.r.t. $\mathbf{F}^L$ and $(\mathbf{F}^{*})^L$
in the last layer as
	\begin{equation}
		\label{G_FL}
		\mathbf{G}_F^L = (\nabla_{\mathbf{F}} \tilde{\mathcal{P}}_e^L)^H,  \quad
		\mathbf{G}_{F^*}^L = (\nabla_{\mathbf{F} ^{*} } \tilde{\mathcal{P}}_e^L)^H. \vspace{-2mm}
	\end{equation}
By differentiating on both sides of \eqref{delta_theu} and \eqref{delta_thef}, it is readily seen that $\mathbf{G}_{\theta_{U_k}}^l$ and $\mathbf{G}_{\theta_F}^l$ can be computed based on $\{\mathbf{G}_{U_k}^l, \mathbf{G}_{{U_k^*}}^l \}$ and $\{\mathbf{G}_F^l, \mathbf{G}_{F^*}^l \}$, respectively, as
\vspace{-0mm}
	\begin{equation}
		\quad \mathbf{G}_{\theta_{U_k}}^l = \mathbf{G}_{U_k}^l \circ j \mathbf{U}_k^T - \mathbf{G}_{U_k^*}^l \circ j  \mathbf{U}_k^*,     \vspace{-0mm}        \label{G_theu}
	\end{equation}
	\begin{equation}
		\mathbf{G}_{\theta_{F}}^l = \mathbf{G}_F^l \circ j  \mathbf{F}^T - \mathbf{G}_{F^*}^l \circ j \mathbf{F}^*.  \vspace{-0mm}      \label{G_thef}
	\end{equation}

Next, we derive the recursive relationship between the gradients
w.r.t. the hybrid beamforming matrices in the ($l+1$)-th layer and the
$l$-th layer.  To this end, we first take the derivative on both sides
of the equations \eqref{update_p_new}--\eqref{update_F_new} and apply
the differential multiplication rules. Let us first expand
\eqref{update_p_new} and provide the relationship between
$\mathbf{G}_{P_k}^{l+1}$ and $\mathbf{G}_{P_k}^{l}$; the corresponding
relations for the other matrices can be derived similarly. To simplify
the presentation, we omit the index of data stream $i$, the index of
transmit vector $q_j$, and the index of iteration $t$.
Based on \eqref{update_p_new} and the GCR shown in Appendix \ref{appendixA},
we have \vspace{-2mm}
	\begin{equation}
	\begin{split}
	&\!\!\!\!\! {\rm Tr} \left\{\mathbf{G}_{P_k}^{l+1} d\mathbf{P}_k^{l+1} \right\} = \frac{1}{J} \sum \limits_{j=1}^J {\rm Tr} \left\{\mathbf{G}_{P_k}^{l+1} d\mathbf{P}_k^{l} - b_k^H B \, \mathbf{G}_{P_k}^{l+1}  \circ
	\right.\\
	& \!\!\!\!\! (\bm{\alpha}_{P_k}^{l})^T \, \left. (\mathbf{F}^{l})^H \mathbf{H}_k^H (\mathbf{U}_k^{l})^H \mathbf{W}^{l}_k \circ \left[b_k \, (\mathbf{W}_k^{l})^H \mathbf{U}_k^{l} \mathbf{H}_k \mathbf{F}^{l}
	\right. \right. \\
	& \!\!\!\!\! d\mathbf{P}_k^{l} + \mathbf{D}^H (\mathbf{U}_k^{l})^H  d\mathbf{W}_k^{l} + \mathbf{C} (\mathbf{W}_k^{l})^H \mathbf{U}_k^{l} \mathbf{H}_k \, d\mathbf{F}^{l} + \mathbf{D}
	\\
	& \!\!\!\!\! \left. \left. (\mathbf{W}_k^{l})^H d\mathbf{U}_k^{l} + \mathbf{H}_k^H (\mathbf{U}_k^{l})^H \mathbf{W}_k^{l} \mathbf{C}^H d(\mathbf{F}^{H})^{l} + \mathbf{W}_k^{l} \mathbf{D}^H
	\right. \right. \\
	& \!\!\!\!\! \left. \left. d(\mathbf{U}_k^{H})^{l} \right] + b_k^H A \,  \mathbf{G}_{P_k}^{l+1} \circ (\bm{\alpha}_{P_k}^{l})^T (\mathbf{F}^{l})^H \mathbf{H}_k^H (\mathbf{U}_k^{l})^H
	\right.\\
	& \!\!\!\!\! \left.  d\mathbf{W}_k^{l} + b_k^H A \, \mathbf{H}_k^H (\mathbf{U}_k^{l})^H \mathbf{W}_k^{l} \mathbf{G}_{P_k}^{l+1} \circ (\bm{\alpha}_{P_k}^{l})^T d(\mathbf{F}^{H})^{l}
	\right.\\
	& \!\!\!\!\! \left.  + b_k^H A \, \mathbf{W}_k^{l} \mathbf{G}_{P_k}^{l+1} \circ (\bm{\alpha}_{P_k}^{l})^T (\mathbf{F}^{l})^H \mathbf{H}_k^H d(\mathbf{U}_k^{H})^{l} \right\},
	\end{split}\vspace{-4mm}
	\end{equation}
where $A \triangleq \frac{ \Re \{ b_k \} } { \sqrt{2\pi} \rho_{P_k}^{l}} \, e^{-\frac{|\Re \{\bar{b}_k\}|^2} {2(\rho_{P_k}^{l})^2}}$,
$B \triangleq A \, \dfrac{\Re \{ b_k \}}{(\rho_{P_k}^{l})^2}$,
$\mathbf{C} \triangleq \sum \limits_{i=1}^{D_k} \mathbf{p}_{i,k}^{l} b_{i,k}$, and
$\mathbf{D} \triangleq \mathbf{H}_k \mathbf{F}^{l} \mathbf{C}$. \vspace{2mm}
By proceeding in the same way with \eqref{update_w_new}--\eqref{update_F_new}, we obtain all the necessary relations linking the various gradients in adjacent layers.

By isolating and rearranging corresponding terms in $d\mathbf{P}_k^l$, we obtain the desired relationship for $\mathbf{G}_{P_k}^{l}$ as \vspace{-2mm}
	\begin{equation}
	\begin{split}
	& \!\!\!\mathbf{G}_{P_k}^{l} = -  \frac{1}{J} \sum \limits_{j=1}^J \Big(b_k^H B \, \mathbf{G}_{P_k}^{l+1} \circ (\bm{\alpha}_{P_k}^{l})^T (\mathbf{F}^{l})^H \mathbf{H}_k^H (\mathbf{U}_k^{l})^H  \Big.
	\\
	& \ \circ \mathbf{W}_k^{l} \mathbf{E} - B \, \mathbf{G}_{W_k}^{l+1} \circ (\bm{\alpha}_{W_k}^{l})^T \mathbf{U}^{l}_k \mathbf{D} \circ \mathbf{E}
	\\
	& \ + b_k A \, \left[\mathbf{G}_{F}^{l+1} \circ (\bm{\alpha}_{\theta_F}^{l})^T \mathbf{H}_k^T (\mathbf{U}_k^{l+1})^T (\mathbf{W}_k^{l})^{H} \right]^T
	\\
	& \ - b_k A \, \mathbf{G}_{W_k}^{l+1} \circ (\bm{\alpha}_{W_k}^{l})^T \mathbf{U}_k^{l} \mathbf{H}_k \mathbf{F}^{l} + b_k A \, [(\mathbf{F}^{l})^T \mathbf{H}_k^T
	\\ \Big.
	& \,\ \mathbf{G}_{U_k}^{l+1} \circ (\bm{\alpha}_{\theta_{U_k}}^{l})^T (\mathbf{W}_k^{l})^{H} ]^T - B \, \mathbf{D}^T \mathbf{G}_{U_k}^{l+1} \circ (\bm{\alpha}_{\theta_{U_k}}^{l})^T
	\\
	& \,\ (\mathbf{W}_k^{l})^{H} \circ \mathbf{E} - B \, \mathbf{C}^H \mathbf{G}_{F^*}^{l+1} \circ (\bm{\alpha}_{\theta_F}^{l})^T \mathbf{H}_k^H (\mathbf{U}_k^{l+1})^H
	\\
	& \,\ \mathbf{W}_k^{l} \circ \mathbf{E}  - B \, \mathbf{D}^H \mathbf{G}_{U_k^*}^{l+1} \circ (\bm{\alpha}_{\theta_{U_k}}^{l})^T \mathbf{W}_k^{l} \circ \mathbf{E} - B \, \mathbf{C}^T
	\\
	& \ \ \mathbf{G}_{F}^{l+1} \circ (\bm{\alpha}_{\theta_F}^{l})^T \mathbf{H}_k^T (\mathbf{U}_k^{l+1})^T (\mathbf{W}_k^{l})^{H} \circ \mathbf{E} \Big) + \mathbf{G}_{P_k}^{l+1} ,
	\end{split}
	\end{equation}
where $\mathbf{E} \triangleq b_k \, (\mathbf{W}_k^{l})^H
\mathbf{U}_k^{l} \mathbf{H}_k \mathbf{F}^{l}$.  The gradients
w.r.t. the other beamforming matrices in the $l$-th layer can be
obtained similarly.  Furthermore, the gradients w.r.t. the introduced
parameters in each layer are computed based on
\eqref{update_p_new}--\eqref{update_F_new}. The detailed expressions
of $\{\nabla_{\bm{\alpha}_X} \tilde{\mathcal{P}}_e^l, \nabla_{\rho_X}
\tilde{\mathcal{P}}_e^l, \nabla_{\mathbf{O}_X} \tilde{\mathcal{P}}_e^l
\}$, where $X \in \{P_k, W_k, \theta_{U_k}, \theta_F\}$, are shown in
Appendix \ref{appendixB}.

We calculate the average gradient in a batch and implement the SGD
method to update the trainable parameters, such as in, e.g.,
$\bm{\alpha}_{P_k}^{l,t+1} = \bm{\alpha}_{P_k}^{l,t} -
\mu_{\alpha_{P_k}}^{t} \nabla_{\bm{\alpha}_{P_k}^{t}}
\tilde{\mathcal{P}}_e^l$, where $\mu_{\alpha_{P_k}}^{t}$ denotes the
step size of the update in SGD of $\bm{\alpha}_{P_k}^{l}$ in the $t$-th iteration, which incorporates an attenuation factor dependent on iteration $t$.
In order to avoid vanishing gradient problem in the BP, normalization is employed after the update of $\{\mathbf{G}_{P_k}^l,
\mathbf{G}_{W_k}^l, \mathbf{G}_{U_k}^l, \mathbf{G}_{U^*_k}^l,
\mathbf{G}_F^l, \mathbf{G}_{F^*}^l \}$. The specific rule of
normalizing $\mathbf{G}_{P_k}^l$ is given as \vspace{-2mm}
	\begin{equation}
		\mathbf{G}_{P_k}^l \leftarrow \dfrac{K \mathbf{G}_{P_k}^l}{\sum \limits_k \! \left\|\mathbf{G}_{P_k}^l\right\| }. \vspace{-2mm}
	\end{equation}
The gradients w.r.t. the other variables can be computed in the same
way. The trainable parameters are initialized randomly and the
beamforming matrices $\{\mathbf{P}_k^0, \mathbf{W}_k^0,
\bm{\theta}_{U_k}^0, \bm{\theta}_{F}^0 \}$ are initialized based on
the conventional channel alignment method \cite{2013Spatially}. The
training process of the deep-unfolding NN is shown in Algorithm
\ref{tab:deep-unfolding}, where $\mathcal{H} \triangleq
\{\mathbf{H}_k^1, \mathbf{H}_k^2,\ldots, \mathbf{H}_k^N \}$ and $N$ is
determined by simulations.

\begin{algorithm}[t]
	\caption{Training process of the proposed deep-unfolding NN induced by GD algorithm}
	\begin{algorithmic}[1]
		\begin{spacing}{0.9}
			\begin{small} \vspace{0mm}
				\State \parbox[t]{\dimexpr\linewidth-\algorithmicindent-\algorithmicindent}{Generate training data set $\{\mathcal{X}, \mathcal{H} \}$. Set tolerance of accuracy $\epsilon$, number of layers $L$, batch size $N$, the maximum number of iterations $I_{\max}$, and the size of the random sample $J$. Set the current iteration index $t=0$. Initialize the beamforming matrices, trainable parameters, and step sizes.} \vspace{2mm}
				\Repeat  \vspace{0.5mm}
				\State \parbox[t]{\dimexpr\linewidth-\algorithmicindent-\algorithmicindent}{\textbf{Forward propagation}: Randomly select $J$ samples $\{\mathbf{s}, \mathbf{H}_k, \forall k \}$ from $\{\mathcal{X}, \mathcal{H}\}$. Calculate $\{\bm{\theta}_F^l, \mathbf{F}^l, l = 1,\ldots,L-1 \}$ and $\{\mathbf{P}_k^l, \mathbf{W}_k^l, \bm{\theta}_{U_k}^l, \mathbf{U}_k^l, l = 1,\ldots,L, \forall k \}$ based on \eqref{update_p_new}--\eqref{update_F_new}.}  \vspace{1mm}
				\State \parbox[t]{\dimexpr\linewidth-\algorithmicindent-\algorithmicindent}{Calculate $\bm{\theta}_{F}^L$ based on \eqref{update_F}. Calculate $\mathbf{F}^L$ based on $\bm{\theta}_{F}^L$. Substitute $\{\mathbf{P}_k^L, \mathbf{W}_k^L, \mathbf{U}_k^L, \mathbf{F}^L \}$ into \eqref{objective function}.}  \vspace{1mm}
				\State \parbox[t]{\dimexpr\linewidth-\algorithmicindent-\algorithmicindent}{\textbf{Backward propagation}: Firstly, calculate the gradients of $\{\mathbf{F}^L, (\mathbf{F}^{L})^{*}\}$ in the last layer and the gradient of $\bm{\theta}_{F}^L$ based on \eqref{G_thef}. Secondly, calculate the gradients of $\{\mathbf{P}_k^l, \mathbf{W}_k^l, \mathbf{U}_k^l, (\mathbf{U}_k^{l})^{*}, \bm{\theta}_{U_k}^l, l = L, \ldots, 0, \forall k \}$ and  $\{\mathbf{F}^l, (\mathbf{F}^{l})^{*}, \bm{\theta}_F^l, l = L-1, \ldots, 0 \}$  based on Appendix \ref{appendixA}. Finally, calculate the gradients of the trainable parameters based on Appendix \ref{appendixB}.}  \vspace{1mm}
				\State \parbox[t]{\dimexpr\linewidth-\algorithmicindent-\algorithmicindent}{Calculate the average gradient in a batch and update the trainable parameters based on the SGD method.} \vspace{1mm}
				\State \parbox[t]{\dimexpr\linewidth-\algorithmicindent-\algorithmicindent}{Update the iteration number : $t=t+1$.}
				\Until{The loss function in the validation data set converges or $t > {I_{\max}}$.}  \vspace{-4mm}
			\end{small}
		\end{spacing}
	\end{algorithmic}              \label{tab:deep-unfolding}
\end{algorithm}

\vspace{-3mm}
\section{Algorithm Analysis}
\label{Section5:analysis}

In this section, we demonstrate that the sequence of iterates generated by the proposed deep-unfolding NN is convergent.
Then we develop a black-box CNN as a benchmark to jointly optimize the hybrid AD beamforming matrices. Moreover, we analyze the computational complexity and generalization ability of the proposed schemes. 
Finally, we extend the deep-unfolding NN to $M$-QAM signal constellations.

\vspace{-4mm}
\subsection{Convergence of deep-unfolding NN}
In general, no claim of guaranteed convergence can be made for existing deep-unfolding NNs due to the introduction of trainable parameters in the deep-unfolding NN as well as structural differences between the latter and the original iterative optimization algorithm.
In the following, we circumvent some of these difficulties and provide novel theoretical insight into the convergence of the deep-unfolding NN.
It is difficult to strictly prove its convergence.
In the following, we provide some theoretical analysis for the convergence of the deep-unfolding NN.

\textit{\textbf{Theorem 1} (Convergence of deep-unfolding NN):
The performance of one layer in the deep-unfolding NN can approach that of several iterations in the GD algorithm if the parameters are properly trained.
Consequently:
\begin{enumerate}[1)]
	\item The performance of several layers in the deep-unfolding NN can approach that of the iterative GD algorithm.
	\item The deep-unfolding NN converges to a statonary point with much reduced number of layers.
\end{enumerate}
}

The proofs of these claim can be provided as follows. As shown in Section \ref{Section3:gradient-descent}, in the $t$-th iteration of the GD algorithm, we have the following mapping from $\mathbf{P}_{t}$ to $\mathbf{P}_{t+2}$, where we omit indices $k$ and $i$ for clarity: \vspace{-2mm}
\begin{equation}  \label{twoiter}
\begin{split}
&\!\!\!\mathbf{P}_{t+2} = \mathbf{P}_{t} - \mu_P A \sum \limits_{{j}=1}^J \, e^{-\frac{|\Re \{\bar{b}^{q_j}\}|^2} {2\varrho^2}} \ \Re \{b\}  \mathbf{F}_{t}^H \mathbf{H}^H \mathbf{U}_{t}^H \mathbf{W}_{t} (\mathbf{b}^{q_j})^H
\\
& - \mu_P A \sum \limits_{{j}=1}^J \, e^{-\frac{|\Re \{\bar{b}^{q_j}\}|^2} {2\varrho^2}} \ \Re \{b\}  \mathbf{F}_{t+1}^H \mathbf{H}^H \mathbf{U}_{t+1}^H \mathbf{W}_{t+1} (\mathbf{b}^{q_j})^H.
\end{split}\vspace{-4mm}
\end{equation}
Similarly, in the deep-unfolding NN, we have the following mapping from $\mathbf{P}_{l}$ to $\mathbf{P}_{l+1}$ in the $l$-th layer: \vspace{-2mm}
\begin{equation} \label{duNN}
\!\!\!\mathbf{P}_{l+1} \!\!=\! \mathbf{P}_{l} - \bm{\alpha}_{P}^l \circ \big(\!A\! \sum \limits_{{j}=1}^J \!\! e^{-\frac{|\Re \{\bar{b}^{q_j}\}|^2} {2(\rho_P^l)^2}} \! \Re \{b\}  \mathbf{F}_{\!l}^{\!H} \mathbf{H}^{\!H} \mathbf{U}_{l}^{\!H} \mathbf{W}_{\!l} (\mathbf{b}^{q_j}\!)^{\!H} \big) \!+\! \mathbf{O}_{P}^l.
\end{equation}
On the basis of these relations, we prove that one layer in the deep-unfolding NN can approach two iterations in the GD algorithm. In detail, when the initial values of the two algorithms are identical, i.e., $\mathbf{P}_{l}=\mathbf{P}_{t}$, we need to prove that $\mathbf{P}_{l+1}$ approaches $\mathbf{P}_{t+2}$, i.e., $\|\mathbf{P}_{t+2} - \mathbf{P}_{l+1}\|^2 < \varepsilon$, for any $\varepsilon>0$. In the following, we provide the analytical justification for two different cases.

\vspace{-0mm}
\subsubsection{\textbf{Case 1} (Fixed, i.e., deterministic channel)} When the channel matrix $\mathbf{H}$ is fixed or only changes very slowly,
we are supposed to demonstrate that there exist trainable parameters $\bm{\alpha}_{X}$, $\rho_{X}$, and $\mathbf{O}_{X}$ such that $\|\mathbf{P}_{t+2} - \mathbf{P}_{l+1}\|^2 < \varepsilon$ is satisfied for a given $\mathbf{H}$. By comparing \eqref{twoiter} with \eqref{duNN}, we can make $\mathbf{P}_{t+2}= \mathbf{P}_{l+1}$ satisfied if we set \vspace{-1mm}
\begin{subequations}
	\begin{eqnarray}
	& & \!\!\!\!\!\!\!\!\bm{\alpha}_P^l = \mu_P \, \mathbf{1}^{R_t \times D_k},  \\
	& & \!\!\!\!\!\!\!\!\mathbf{O}_P^l = - \mu_P A \sum \limits_{{j}=1}^J \, e^{-\frac{|\Re \{\bar{b}^{q_j}\}|^2} {2\varrho^2}} \ \Re \{b\}  \mathbf{F}_{t+1}^H \mathbf{H}^H \\ \vspace{-2mm}
	& & \!\!\!\!\!\!\!\!\qquad \quad \mathbf{U}_{t+1}^H \mathbf{W}_{t+1} (\mathbf{b}^{q_j})^H, \notag \\
	& & \!\!\!\!\!\!\!\!\rho_P^l = \varrho,
	\end{eqnarray}
\end{subequations}
where $\mathbf{1}^{R_t \times D_k}$ denotes the matrix with dimension $R_t \times D_k$ and all elements equal to $1$.

\vspace{-0mm}
\subsubsection{\textbf{Case 2} (Channel Following Certain Distribution)} When the channel matrix $\mathbf{H}$ conforms to a
certain distribution, we have to illustrate the existence of parameters $\bm{\alpha}_{X}$, $\rho_{X}$, and $\mathbf{O}_{X}$ such that the following inequality is satisfied:
\begin{equation}
\mathbb{E}_{\mathbf{H}}\Big\{\|\mathbf{P}_{t+2} - \mathbf{P}_{l+1}\|^2 \Big\} \leq \varepsilon.
\end{equation}
Setting $\bm{\alpha}_P^l = \mu_P \, \mathbf{1}^{R_t \times D_k}$ and $\rho_P^l = \varrho$, we need to prove
\begin{equation}  \label{final problem}
\!\!\!\mathbb{E}_{\mathbf{H}} \! \Big\{\!\|\mathbf{O}_P - \mu_P A \! \sum \limits_{{j}=1}^J \!\! e^{\!-\!\frac{|\!\Re \{\bar{b}^{\!q_{\!j}} \!\} \!|^2} {2\varrho^2}} \Re\{b\}  \mathbf{F}_{\!t\!+\!1}^{\!H} \mathbf{H}^{\!H} \mathbf{U}_{\!t\!+\!1}^{\!H} \mathbf{W}_{\!t\!+\!1} (\mathbf{b}^{\!q_{\!j}}\!)^{\!H} \!\|^{\!2}\!\Big\} \!\leq\! \varepsilon.
\end{equation}
The variables $\mathbf{F}_{t+1}$, $\mathbf{U}_{t+1}$, and $\mathbf{W}_{t+1}$ are all related to $\mathbf{P}_t$. To simplify the presentation, we only expand $\mathbf{W}_{t+1}$ here but the other variables can be handled similarly.
The left side of \eqref{final problem} can be expressed as \vspace{-0mm}
\begin{spacing}{0.1}
\begin{equation}  \label{proof}
\begin{split}
&\!\!\!\mathbb{E}_{\mathbf{H}} \! \Big\{\|\mathbf{O}_P - \mu_P A \! \sum \limits_{{j}=1}^J \!\! e^{\!-\!\frac{|\!\Re \{\bar{b}^{\!q_{\!j}} \!\} \!|^2} {2\varrho^2}} \Re\{b\}  \mathbf{F}_{\!t\!+\!1}^{\!H} \mathbf{H}^{\!H} \mathbf{U}_{\!t\!+\!1}^{\!H} \mathbf{W}_{\!t\!+\!1} (\mathbf{b}^{q_{j}}\!)^{\!H} \|^{2}\Big\}
\\
&\!\!\!\!\! = \mathbb{E}_{\mathbf{H}} \! \Big\{\|\mathbf{O}_P - \mu_P A \! \sum \limits_{{j}=1}^J \!\! e^{\!-\!\frac{|\!\Re \{\bar{b}^{\!q_{\!j}} \!\} \!|^2} {2\varrho^2}} \Re\{b\}  \mathbf{F}_{\!t\!+\!1}^{\!H} \mathbf{H}^{\!H} \mathbf{U}_{\!t\!+\!1}^{\!H} \mathbf{W}_{\!t} (\mathbf{b}^{q_{j}}\!)^{\!H} \|^{2}\Big\}
\\
&\!\!\!\! +\! \mu_P \mu_W \!A^2\! \sum \limits_{{j}=1}^J \!\sum \limits_{{m}=1}^J \!\! e^{-\frac{|\!\Re \{\bar{b}^{\!q_{\!j}} \!\} \!|^2 + |\!\Re \{\bar{b}^{\!q_{\!m}} \!\} \!|^2} {2\varrho^2}} \mathbf{F}_{\!t\!+\!1}^{\!H} \mathbf{H}^{\!H} \mathbf{U}_{\!t\!+\!1}^{\!H} \mathbf{U}_{\!t} \mathbf{H} \mathbf{F}_{\!t} \!\! \sum \limits_{k'=1}^K \! \big(\mathbf{P}_{\!t}
\\
&\!\!\mathbf{b}^{q_m} (\mathbf{b}^{q_j}\!)^{\!H}\big) \!- \!\mu_P^2 \mu_W \!A^3\! \sum \limits_{{j}=1}^J \!\sum \limits_{{m}=1}^J \!\sum \limits_{{n}=1}^J \!\! e^{-\frac{|\!\Re \{\bar{b}^{\!q_{\!j}} \!\} \!|^2 + |\!\Re \{\bar{b}^{\!q_{\!m}} \!\} \!|^2 + |\!\Re \{\bar{b}^{\!q_{\!n}} \!\} \!|^2} {2\varrho^2}} \! \Re\{ b \}
\\
&\!\! \mathbf{F}_{\!t\!+\!1}^{\!H} \mathbf{H}^{\!H} \mathbf{U}_{\!t\!+\!1}^{\!H} \mathbf{U}_{\!t} \mathbf{H} \mathbf{F}_{\!t} \mathbf{F}_{\!t}^{\!H} \mathbf{H}^{\!H} \mathbf{U}_{t}^H \mathbf{W}_{t} (\mathbf{b}^{q_n}\!)^{\!H} \sum \limits_{k'=1}^K \mathbf{b}^{q_m} (\mathbf{b}^{q_j}\!)^{\!H} \|^2\Big\}
\\
&\!\!\!\!\! \leq \mathbb{E}_{\mathbf{H}} \! \Big\{\|\mathbf{O}_P - \mu_P A \! \sum \limits_{{j}=1}^J \!\! e^{\!-\!\frac{|\!\Re \{\bar{b}^{\!q_{\!j}} \!\} \!|^2} {2\varrho^2}} \Re\{b\}  \mathbf{F}_{\!t\!+\!1}^{\!H} \mathbf{H}^{\!H} \mathbf{U}_{\!t\!+\!1}^{\!H} \mathbf{W}_{\!t} (\mathbf{b}^{q_{j}}\!)^{\!H} \|^{2}\Big\}
\\
&\!\!\!\! +\! \mathbb{E}_{\mathbf{H}} \Big\{\!\sum \limits_{{j}=1}^J \!\sum \limits_{{m}=1}^J\! \sum \limits_{k'=1}^K \!\!\| \mu_P \mu_W \!A^2 \! e^{-\frac{|\!\Re \{\bar{b}^{\!q_{\!j}} \!\} \!|^2 + |\!\Re \{\bar{b}^{\!q_{\!m}} \!\} \!|^2} {2\varrho^2}}\|^2 \|\mathbf{F}_{\!t\!+\!1}^{\!H} \mathbf{H}^{\!H} \mathbf{U}_{\!t\!+\!1}^{\!H} \|^2
\\
&\!\!\! \|\mathbf{U}_{\!t} \mathbf{H} \mathbf{F}_{\!t}\|^2 \|\mathbf{P}_{\!t}\mathbf{b}^{q_m} (\mathbf{b}^{q_j}\!)^{\!H} \|^2\Big\} \!+\! \mathbb{E}_{\mathbf{H}} \Big\{\sum \limits_{{j}=1}^J \!\sum \limits_{{m}=1}^J \!\sum \limits_{{n}=1}^J \!\sum \limits_{k'=1}^K \!\!|| \mu_P^2 \mu_W \!A^3 \!
\\
&\!\!\! e^{-\frac{|\!\Re \{\bar{b}^{\!q_{\!j}} \!\} \!|^2 + |\!\Re \{\bar{b}^{\!q_{\!m}} \!\} \!|^2 + |\!\Re \{\bar{b}^{\!q_{\!n}} \!\} \!|^2} {2\varrho^2}}||^2 \|\mathbf{F}_{\!t\!+\!1}^{\!H} \mathbf{H}^{\!H} \mathbf{U}_{\!t\!+\!1}^{\!H} \|^2 \|\mathbf{U}_{\!t} \mathbf{H} \mathbf{F}_{\!t} \|^4
\\
&\!\!\! \|\mathbf{W}_{\!t} (\mathbf{b}^{q_n}\!)^{\!H} \mathbf{b}^{q_m} (\mathbf{b}^{q_j}\!)^{\!H} \|^2\Big\},  \vspace{-0mm}
\end{split}
\end{equation}
\end{spacing}
\noindent where the first equality is obtained based on \eqref{update_w} and \eqref{delta_W}, and the second inequality is derived based on \textit{The Absolute Value Inequality}.
The term $\mu_P A \sum \limits_{{j}=1}^J \, e^{-\frac{|\Re \{\bar{b}^{q_j}\}|^2} {2\varrho^2}} \ \Re \{b\}  \mathbf{F}_{t+1}^H \mathbf{H}^H \mathbf{U}_{t+1}^H \mathbf{W}_{t} (\mathbf{b}^{q_j})^H$ is a function of $\mathbf{H}$ and its mean value, denoted as $\gamma$ can be calculated based on $\mathbb{E}\{\mathbf{H}\}$. According to \textit{The Law of Large Numbers}, when we sample enough $\mathbf{H}$ in the calculation,
as long as $\mathbf{O}_P$ is set to $\gamma$, \eqref{proof} can be simplified as \vspace{-2mm}
\begin{equation}   \label{proof_final}
\begin{split}
&\mathbb{E}_{\mathbf{H}} \Big\{||\mathbf{P}_{\!t\!+\!2\!} - \mathbf{P}_{\!l\!+\!1\!} ||^2\Big\}\leq \mathbb{E}_{\mathbf{H}} \Big\{\!\sum \limits_{{j}=1}^J \!\sum \limits_{{m}=1}^J \!\sum \limits_{k'=1}^K \!\! \| \mu_P \mu_W \!A^2 \!
\\
& e^{-\frac{|\!\Re \{\bar{b}^{\!q_{\!j}} \!\} \!|^2 + |\!\Re \{\bar{b}^{\!q_{\!m}} \!\} \!|^2} {2\varrho^2}}\|^2 \|\mathbf{F}_{\!t\!+\!1}^{\!H} \mathbf{H}^{\!H} \mathbf{U}_{\!t\!+\!1}^{\!H} \|^2 \|\mathbf{U}_{\!t} \mathbf{H} \mathbf{F}_{\!t}\|^2 \|\mathbf{P}_{\!t}\mathbf{b}^{q_m} (\mathbf{b}^{q_j}\!)^{\!H} \|^2\Big\}
\\
& \!\!+ \mathbb{E}_{\mathbf{H}} \Big\{\sum \limits_{{j}=1}^J \!\sum \limits_{{m}=1}^J \!\sum \limits_{{n}=1}^J \!\sum \limits_{k'=1}^K \!\! \|\mu_P^2 \mu_W \!A^3 \! e^{-\frac{|\!\Re \{\bar{b}^{\!q_{\!j}} \!\} \!|^2 + |\!\Re \{\bar{b}^{\!q_{\!m}} \!\} \!|^2 + |\!\Re \{\bar{b}^{\!q_{\!n}} \!\} \!|^2} {2\varrho^2}}\|^2 \
\\
& \!\|\mathbf{F}_{\!t\!+\!1}^{\!H} \mathbf{H}^{\!H} \mathbf{U}_{\!t\!+\!1}^{\!H} \|^2 \|\mathbf{U}_{\!t} \mathbf{H} \mathbf{F}_{\!t} \|^4 \|\mathbf{W}_{\!t} (\mathbf{b}^{q_n}\!)^{\!H} \mathbf{b}^{q_m} (\mathbf{b}^{q_j}\!)^{\!H} \|^2\Big\}  =  \varepsilon.\vspace{-0mm}
\end{split}
\end{equation}
Through the experiment, we can see that the product terms in the right side of \eqref{proof_final} are all less than $1$, so $\varepsilon$ is a term which is far less than $1$. Indeed, $\varepsilon$ is a loose boundary between the performance of the deep-unfolding NN and the iterative GD algorithm.

In the above, we have proved that the performance of one layer in the deep-unfolding NN can approach that of two iterations in the GD algorithm.
It then naturally follows that the performance of one layer in the deep-unfolding NN with more complex structure can approach that of several iterations in the iterative GD algorithm. Thus, the proposed deep-unfolding NN can approach the GD algorithm.
Moreover, since the GD algorithm is known to converge to a stationary point of its objective function, it follows that the proposed deep-unfolding NN also converges to a stationary point.


\vspace{-2mm}
\subsection{Benchmark CNN}
\label{blackbox}
\vspace{-0mm}

\begin{figure}[t]
	\centering
	\includegraphics[width=0.45\textwidth]{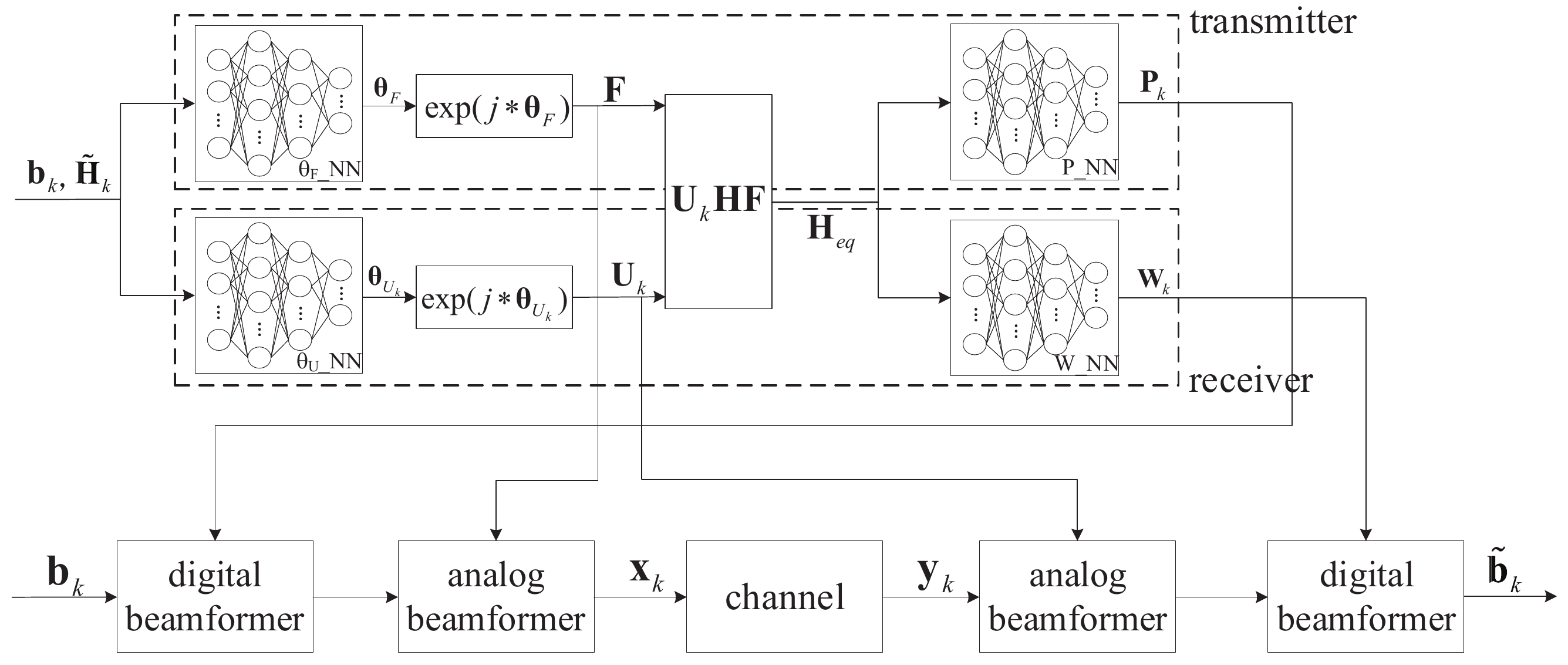}
	\caption{Structure of the benchmark CNN for transceiver design.}
	\label{CNN system framework}
\end{figure}

Recently, the use of NNs to emulate the end-to-end signal transmission in
communication systems has drawn considerable attention \cite{DLendtoend, CNNLi, DLchannelestimation, MLP, CNN, DNN}. In this work, based on the NN structure developed in \cite{CNNLi}, we employ a composite structure comprised of $4$ CNNs for the hybrid AD transceiver design, as shown in Fig. \ref{CNN system framework}. Specifically, the two CNNs labeled ``P\_NN'' and
``$\theta_F$\_NN'' in the top dotted rectangle at the transmitter,
implement the digital transmit beamforming matrices $\mathbf{P}_k$ and
the analog transmit beamforming phase matrix $\bm{\theta}_F$,
respectively, where the original symbol vectors $\mathbf{b}_k$ are
converted into the precoded signals. The CNNs labeled ``W\_NN'' and
``$\theta_U$\_NN'' in the bottom dotted rectangle at the receiver
implement the digital receive beamforming matrices $\mathbf{W}_k$ and
the analog receive beamforming phase matrices $\bm{\theta}_{U_k}$,
respectively, which use the received signals $\mathbf{y}_k$ to
generate $\tilde{\mathbf{b}}_k$. In the training stage, the SER is
obtained by comparing the original symbol vector $\mathbf{b}_k$ with
the detected symbol vector $\hat{\mathbf{b}}_k$ and the SER function
\eqref{eq:mserobjective} is defined as the loss function.  Since
standard CNN implementations (e.g. Tensorflow) cannot process
complex-number matrices directly, the channel matrix $\mathbf{H}_k \in
\mathbb{C}^{N_r \times N_t}$ is transformed into a $2 \! \times \! N_r
\! \times \! N_t$ dimensional real-number tensor
$\tilde{\mathbf{H}}_k$. The inputs of ``$\theta_F$\_NN'' and
``$\theta_U$\_NN'' are $\tilde{\mathbf{H}}_k$ and $\mathbf{b}_k$, and
the outputs are used to generate a low-dimensional equivalent channel
$\mathbf{H}_{eq}$, which are served as the inputs of ``P\_NN'' and
``W\_NN''.

Each CNN consists of the convolutional, pooling, fully-connected and
batch normalization layers, the latter being implemented to avoid the
gradient explosion. To reach a tradeoff between the training overhead
and system performance, we employ a relatively simple network
structure, where the number of layers of each CNN is set to $17$. In
order to satisfy the transmit power constraint, the output of the
``P\_NN'' needs to be normalized in the same way as in
\eqref{scaling}.

In the training stage, the adjustable weight parameters of the CNNs
are updated by the SGD method (available in Tensorflow) while in the
testing stage, only the FP is involved.

\vspace{-3mm}
\subsection{Dimension of Parameter Space}
We first discuss the parameter space dimension of the proposed
deep-unfolding NN followed by the benchmark CNN.  The number of the
parameters introduced in each layer of the deep-unfolding NN is
$2(3K+K N_t D + K N_{r,k} D + K R_{r,k} N_{r,k} + R_t N_t)$. Since the
parameters $\{\bm{\alpha}_{\theta_F}^L, \rho_F^L, \rho_{F^H}^L,
\mathbf{O}_{\theta_F}^L \}$ for the analog transmit beamforming matrix
$\mathbf{F}$ are not used in the last layer, i.e., $L$-th layer, the
total number of parameter dimension in the network is $2KL(3+N_t D +
N_{r,k} D + R_{r,k} N_{r,k}) + 2(L-1)R_t N_t$.

In the benchmark CNN, model parameters are needed to implement the
interconnection of the convolutional layers and the fully-connected
layers. The total number of parameters involved in the convolutional
layers is given by $\sum \limits_{l=1}^{L_c} K_l C_{l-1} C_l$, where
$L_c$ denotes the number of convolutional layers, $K_l$ denotes the
size of convolution kernel, and $C_l$ denotes the number of paths in
the $l$-th convolutional layer. The parameter dimension of the
fully-connected layers is given by $K^2 N_t N_{r,k} F_{in} F_{out}$,
where $F_{in}$ and $F_{out}$ denote the input and output sizes of the
fully-connected layers, respectively.

\vspace{-3mm}
\subsection{Computational Complexity}
The computational complexity of the MSER-based iterative GD algorithm
is given by $\mathcal{O} (TJS (KN_t D + K N_{r,k} D + K R_{r,k}
N_{r,k} + R_t N_t))$, where $T$ denotes the number of iterations and
$S$ denotes the size of the testing data set. As for the black-box
CNN, the computational complexity is given by $\mathcal{O}(\sum
\limits_{l=2}^{L_c} L_c^2 K_l C_{l-1} C_l + K^2 N_t N_{r,k} F_{in}
F_{out} + K N_t D + K N_{r,k} D + K R_{r,k} N_{r,k} + R_t N_t)$, where
$L_c \triangleq (C_{in}-K_l+2P)/S_t + 1$, $C_{in}$ denotes the input
size of the convolutional layers, $P$ denotes the padding size, and
$S_t$ denotes the stride. The computational complexity of the proposed
deep-unfolding NN in the testing stage is $\mathcal{O}(L JS (K N_t
D + K N_{r,k} D + K R_{r,k} N_{r,k} + R_t N_t))$, where $L$ is the
number of layers. In practice, for a comparable level of performance
as shown in Section \ref{Section7:simulations}, we find $L \ll T$;
which means that the deep-unfolding NN can effectively reduce the
computational complexity compared to its iterative GD counterpart.
Also, the parameter space dimension and computational complexity of
the deep-unfolding NN is greatly reduced compared to the benchmark
CNN.

\vspace{-3mm}
\subsection{Generalization Capability}
If a large-scale deep-unfolding NN has been trained, it can be
employed to implement a network with smaller values of $(N_t, N_{r,k},
K)$ directly instead of training a new one. Suppose that the original
large-scale NN is trained with $(N_{t0}, N_{r,k0}, K_0)$ and the
smaller system is characterized by $(N_{t1}, N_{r,k1}, K_1)$, where
$N_{t0} \leq N_{t1}$, $N_{r,k0} \leq N_{r,k1}$, and $K_0 \leq K_1$. In
the testing stage, we only need to input $\{\mathbf{H}_k, k \leq K_1
\}$ and set $\{\mathbf{H}_k = \mathbf{0}, K_1 < k \leq
K_0\}$. Meanwhile, the corresponding column and row vectors in
$\mathbf{H}_k$ should be set to $\mathbf{0}$.

\vspace{-3mm}
\subsection{Extension to  $M$-QAM Modulation}
\label{Section6: QAM}

The proposed deep-unfolding NN can also be extended to  the higher-order modulation scheme, i.e., the $M$-QAM signals.
In this case, the symbol is detected as \vspace{-1mm}
	\begin{equation}    \label{s_QAM}
	\begin{split}
		& \Re\{\hat{b}_{i,k}\} =
		\begin{cases}
		F_1, \quad \ \, \textrm{if} \ \Re\{\tilde{b}_{i,k} \} \leq c_{i,k}(F_1+1) \\
		F_m, \quad \,  \textrm{if} \ c_{i,k}(F_m-1) < \Re\{\tilde{b}_{i,k} \} \leq c_{i,k}(F_m+1),  \\
		F_{\sqrt{M}}, \  \textrm{if} \ \Re\{\tilde{b}_{i,k} \} > c_{i,k}(F_{\sqrt{M}-1}),
		\end{cases}
		\\
		& \Im\{\hat{b}_{i,k}\} =
		\begin{cases}
		F_1, \quad \ \, \textrm{if} \ \Im\{\tilde{b}_{i,k} \} \leq c_{i,k}(F_1+1) \\
		F_n, \quad \ \textrm{if} \ c_{i,k}(F_n-1) < \Im\{\tilde{b}_{i,k} \} \leq c_{i,k}(F_n+1),  \\
		F_{\sqrt{M}}, \ \textrm{if} \ \Im\{\tilde{b}_{i,k} \} > c_{i,k}(F_{\sqrt{M}-1}),
		\end{cases}
	\end{split}
	\end{equation}
where $2 \leq m,n \leq \sqrt{M}-1$ and $c_{i,k} \triangleq \mathbf{w}^H_{i,k} \mathbf{U}_k \mathbf{H}_k \mathbf{F} \mathbf{p}_{i,k}$. Generally, $c_{i,k}$ is a complex value and hence, the following phase rotations are utilized to guarantee that $c_{i,k}$ is real and positive:
\vspace{-2mm}
	\begin{equation}    \label{rotation}
		\begin{split}
			&\mathbf{p}_{i,k} \leftarrow \frac{c_{i,k}}{|c_{i,k}|} \mathbf{p}_{i,k}, \ \ \mathbf{w}_{i,k} \leftarrow \frac{c_{i,k}}{|c_{i,k}|} \mathbf{w}_{i,k},
			\\
			&\mathbf{U}_{k} \leftarrow \frac{c_{i,k}}{|c_{i,k}|} \mathbf{U}_{k},  \ \quad  \mathbf{F} \leftarrow \frac{c_{i,k}}{|c_{i,k}|} \mathbf{F},
		\end{split} \vspace{-2mm}
	\end{equation}
which make the detection rule \eqref{s_QAM} effective.

The number of the legitimate sequences of the transmitted signal
$\mathbf{s}$ in the case of $M$-QAM modulation is $N_b =
M^{D-1}$. Hence, similar to the case of QPSK modulation, we choose $J$
different transmit symbol vectors and define the noise-free part of
$\tilde{b}_{i,k}$ as $\bar{b}_{i,k}$. We then obtain the following
expressions for the SER,
\vspace{-2mm}
	\begin{equation}
		\mathcal{P}_e^R = \frac{\varphi}{J \sqrt{\pi}} \sum \limits_{{j}=1}^J \int_{- \infty}^{-\frac{c_{i,k} \left(\Re \left\{b_{i,k} \right\} - 1 \right) - \Re \left\{\bar{b}^{q_j}_{i,k} \right\} } {\sqrt{2}\varrho}} \ e^{-s^2} \ ds,  \vspace{-3mm}
	\end{equation}
	\begin{equation}
		\mathcal{P}_e^I = \frac{\varphi}{J \sqrt{\pi}} \sum \limits_{{j}=1}^J \int_{- \infty}^{-\frac{c_{i,k} \left(\Im \left\{b_{i,k} \right\} - 1 \right) - \Im \left\{\bar{b}^{q_j}_{i,k} \right\} } {\sqrt{2}\varrho}} \ e^{-s^2} \ ds,
	\end{equation}
where $\varphi \triangleq \frac{2 \sqrt{M} - 2}{\sqrt{M}}$ \cite{MSERQAM}.

The derivation of the MSER-based GD iterative algorithm for the
$M$-QAM modulation follows similar steps as for the QPSK modulation,
leading to similar updates as in \eqref{update_p}--\eqref{update_F}
but where the expressions of the gradient vectors are modified
accordingly. The deep-unfolding NN can be also developed in the same
way, where the parameters introduced have the same function as in the
QPSK case. The details of the two schemes are omitted for brevity. The
feasibility of the deep-unfolding NN for the $M$-QAM signals is
verified by the simulation results in Section
\ref{Section7:simulations}.

\vspace{-2mm}
\section{ Simulation Results}
\label{Section7:simulations}

In this section, we investigate the performance of the proposed
MSER-based GD and deep-unfolding NN algorithms. We first present the
simulation methodology, followed by the investigation of the
convergence in training of the deep-unfolding NN. We then present the
comparative performance results of these algorithms to the benchmark
CNN and other approaches from the literature for both QPSK and $M$-QAM
constellations in the context of massive MIMO transmissions.

\vspace{-3mm}
\subsection{Methodology}
We consider downlink transmission in a multi-user MIMO system as
illustrated in Fig. \ref{fig:structure}. The BS is equipped with $N_t
= 64$ transmit antenna elements and $R_t = 8$ RF chains. We consider
$K = 2$ users, each of which is equipped with $N_{r,k} = 8$ receive
antennas and $R_{r,k} = 4$ RF chains. The BS transmits $D_k=3$ data
streams for each user, i.e., $\forall k \in \mathcal{K}$, consisting
of either QPSK or $M$-QAM symbols. The channel matrix between the BS
and the users is generated according to the model presented in
\eqref{H}.

The data set for training the deep unfolding NN is obtained by
generating $500$ channel matrices and using the transmitted and
received symbols as true and target data. In this stage, the average
of the loss function in \eqref{eq:mserobjective} is used to
approximate the expectation in \eqref{objective function}. We set the
batch size as $N = 20$ and the number of layers as $L = 15$ in the
proposed deep-unfolding NN.

For the comparative performance evaluation, we generate $5000$ channel
matrices from which we obtain the testing data. In the case of the GD
algorithm, we run the GD algorithm $50$ times with randomly selected
initial values and retain the best result as its performance, which
can be treated as an upper bound. We consider the following algorithms
for comparison:

$\bullet$ \textbf{GD}: The beamforming matrices are alternately optimized based on the GD algorithm, as developed  in Section \ref{Section3:gradient-descent}.

$\bullet$ \textbf{OMP}: The near-optimal hybrid beamforming matrices are designed to approximate the optimal unconstrained beamforming matrices based on the OMP method in \cite{2013Spatially}.

$\bullet$ \textbf{MO}: The hybrid beamforming matrices are alternately optimized based on the MO method in \cite{2016Alternating}.

$\bullet$ \textbf{Benchmark CNN}: {
	The hybrid beamforming matrices are optimized jointly by the proposed black-box  CNN presented in Section \ref{blackbox} which is designed based on \cite{CNNLi}}.

$\bullet$ \textbf{Proposed deep-unfolding NN}: The hybrid beamforming matrices are optimized jointly based on the  proposed deep-unfolding NN developed in Section \ref{Section4:deep-unfolding}.

\vspace{-3mm}
\subsection{Convergence of Deep Unfolding NN}

\begin{figure}[t]
	\centering
	\includegraphics[width=0.35\textwidth]{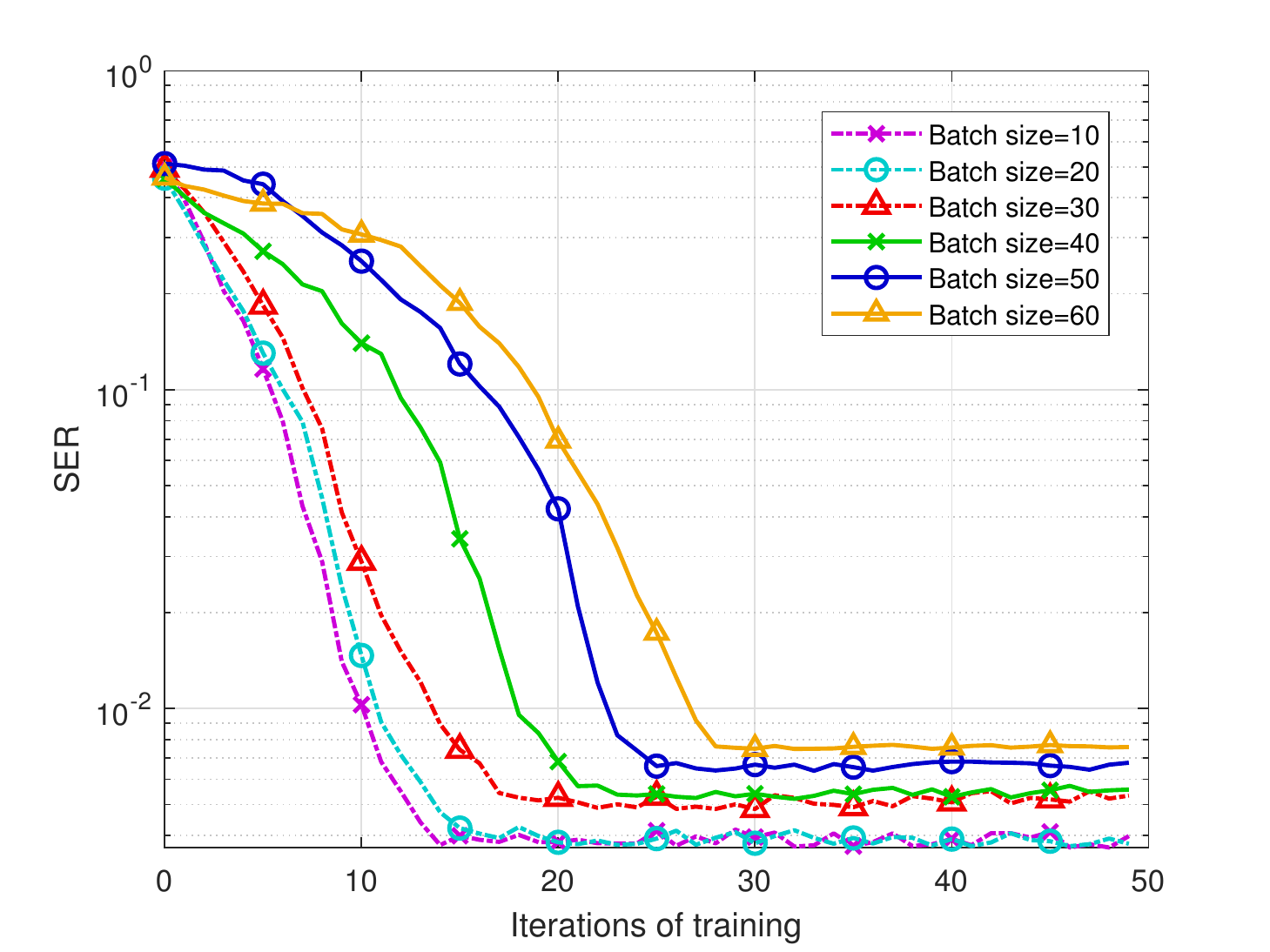}
	\caption{SER of deep-unfolding NN versus iteration number for different batch sizes.}
	\label{result2}
\end{figure}

\begin{figure}[t]
	\centering
	\includegraphics[width=0.35\textwidth]{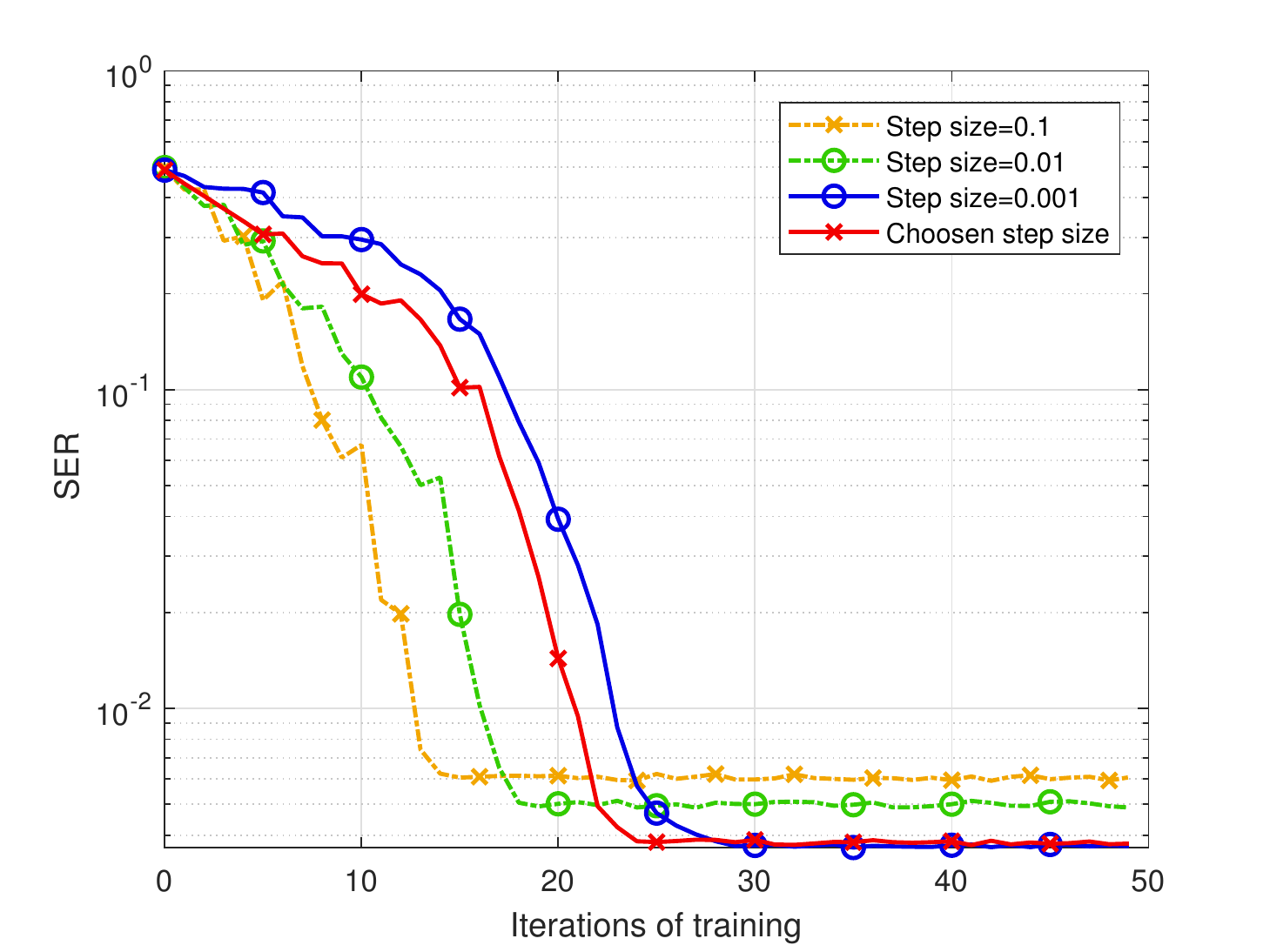}
	\caption{SER of deep-unfolding NN versus iteration number for different step sizes.}
	\label{result3}
\end{figure}

\begin{figure}[t]
	\centering
	\includegraphics[width=0.35\textwidth]{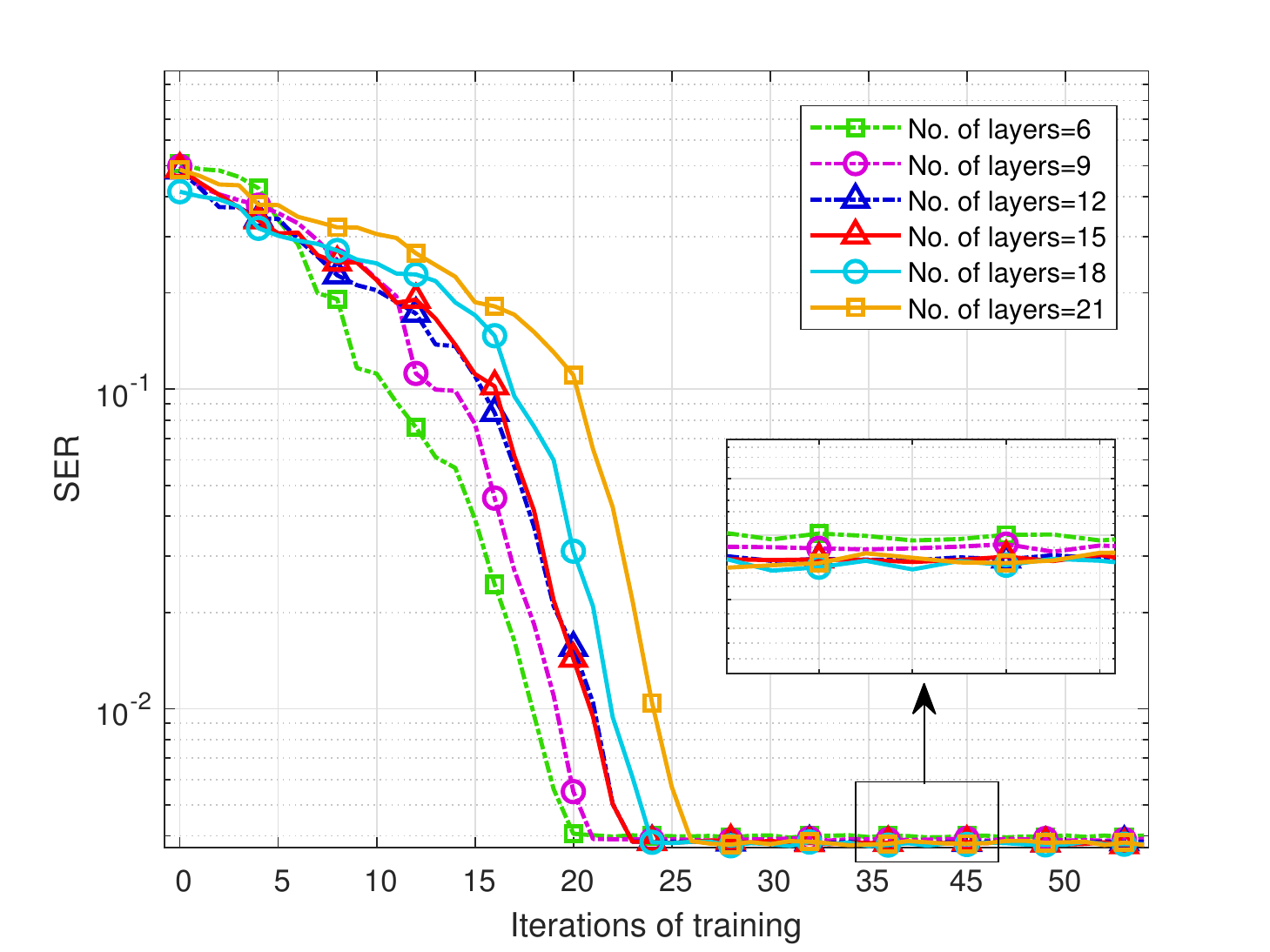}
	\caption{SER of deep unfolding NN versus different number for layers.}
	\label{result4}
\end{figure}

Fig. \ref{result2} illustrates the effects of the batch size on the
convergence speed of the SER for the proposed deep-unfolding NN when
SNR = $20$ dB. As shown, a smaller batch size yields a better SER
performance, in both terms of convergence speed and residual SER after
convergence, while a larger batch size has a smaller fluctuation of
SER performance after convergence. Indeed, a reduction of the batch
size introduces additional randomness in the gradient vector which
prevents the NN from being trapped in local optima.  To provide a good
tradeoff, we choose $N=20$ as the batch size in the sequel.
Fig. \ref{result3} shows the effects of the step size used in the
deep-unfolding NN, i.e., $\{\mu_{\alpha_X}, \mu_{\rho_X},
\mu_{O_X}\}$, on the SER performance when SNR = $20$ dB. The
deep-unfolding NN with a larger step size requires fewer iterations to
converge, but this comes at the price of increased residual SER after
convergence. In this work, we choose the step size as $0.02 \times
0.5^{\lfloor\frac{t}{10}\rfloor}$ which offers a good tradeoff between
system performance and convergence speed.
Fig. \ref{result4} presents the effects of the number of layers $L$ on
the SER performance of the deep-unfolding NN. Increasing the number of
layers slows down convergence (as expected) but also affects the
residual SER after convergence.  As $L$ is increased from $6$ to $18$,
the residual SER decreases until it reaches a lower bound but then
starts to increase.  Indeed, as $L$ is further increased, the
propagation of gradient errors and the effect of gradient
disappearance also become more pronounced. \vspace{-0mm}


\vspace{-3mm}
\subsection{SER Performance for QPSK Signals}

\begin{figure}[t]
	\centering
	\includegraphics[width=0.35\textwidth]{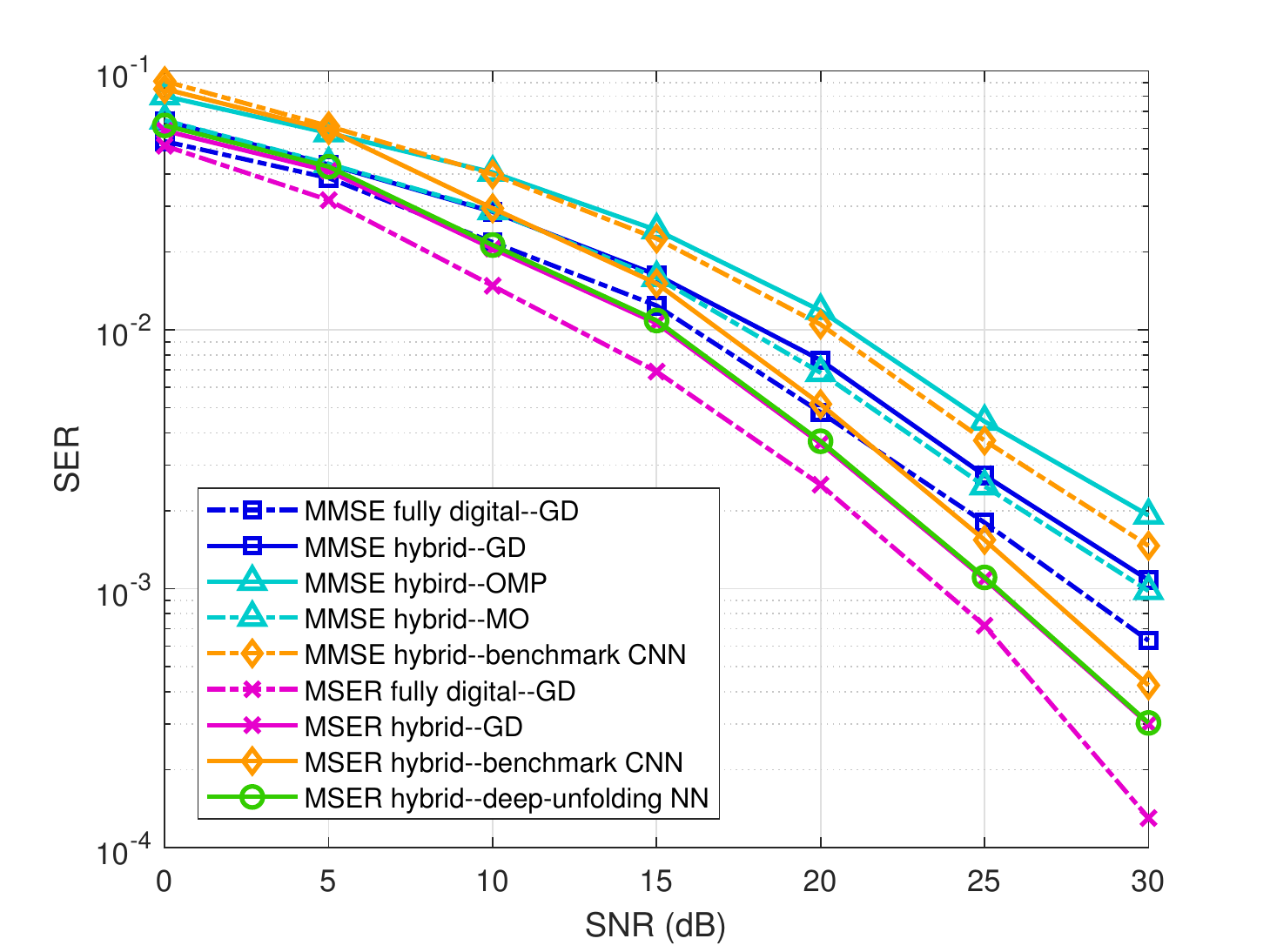}
	\caption{{
			SER performance versus SNR for the different algorithms ($R_{r,k}=4$, $D=3$)}.}
	\label{result1}
\end{figure}

Fig. \ref{result1} shows the SER performance versus SNR in downlink
MIMO transmission for the following schemes: MMSE-based fully digital
GD beamforming algorithm, MMSE-based hybrid GD beamforming algorithm,
MMSE-based hybrid OMP beamforming algorithm, MMSE-based hybrid MO
beamforming algorithm, MSER-based fully digital GD beamforming
algorithm, proposed MSER-based hybrid GD beamforming algorithm, MMSE-based benchmark CNN, MSER-based benchmark CNN and proposed MSER-based deep-unfolding NN. The results show that the proposed MSER-based algorithms significantly outperform the existing MMSE-based algorithms in terms of SER. Moreover, the performance of the proposed deep-unfolding NN approaches that of the MSER-based hybrid GD algorithm, and the gap between their performance decreases with the increase of SNR. The proposed deep-unfolding NN clearly outperforms the benchmark approach; while both algorithms provide much better performance compared to the MMSE-based fully digital algorithm at high SNR.

 Table \ref{table1} presents the SER versus SNR for the proposed
MSER-based deep-unfolding NN, hybrid GD beamforming algorithm and
benchmark CNN, where the number of iterations/layers is emphasized for
comparison. We find that the proposed deep-unfolding NN requires a
smaller number of layers than the black-box CNN to achieve a similar
(or better) performance, while a much larger number of iterations is
required by the GD algorithm to reach that performance
level. Consequently, in the testing stage, the two NN schemes exhibit
a much reduced computational complexity compared to the iterative GD
scheme (see below).
\begin{table}[t]
	\centering
	\caption{SER versus SNR.}         \label{table1}
	\begin{small}
		\renewcommand{\arraystretch}{1.4}
		\setlength{\tabcolsep}{0.5mm}{
			\begin{tabular}{c|c|c|c}
				\toprule  
				\multirow{2}{*}{} & Hybrid GD & Benchmark CNN & Deep-unfolding NN \\
				& $500$ iterations & $17$ layers & $15$ layers \\
				\hline  
				SNR = $5$ dB  & $4.11 \times 10^{-2}$ & $5.92 \times 10^{-2}$ & $4.27 \times 10^{-2}$   \\
				\hline  
				\ SNR = $15$ dB  & $1.06 \times 10^{-2}$ & $1.51 \times 10^{-2}$ & $1.09 \times 10^{-2}$   \\
				\hline  
				\ SNR = $25$ dB  & $1.09 \times 10^{-3}$ & $1.54 \times 10^{-3}$ & $1.11 \times 10^{-3}$   \\
				\bottomrule  
			\end{tabular}}
	\end{small}
\end{table}

Table~\ref{table3} shows the SER for different configurations of the
MIMO system, i.e., different number of users $K$ and number of
receiving RF chains $R_{r,k}$, in the case of SNR = $20$ dB. The
percentages of the deep-unfolding NN in the table are calculated via
dividing the SER of the iterative GD algorithm by those of the
deep-unfolding NN (the percentages in the tables below are the
same). The percentages of the benchmark CNN are calculated in the same
way.  We observe that the performance gap between the proposed
deep-unfolding NN and the MSER-based GD algorithm slightly increases
with $K$. Moreover, the gap between the SER of the general benchmark
CNN and that of the deep-unfolding NN becomes larger with $K$. This
could be the result of increased multi-user interference, which
hinders the ability of the NN to learn from the training
data. However, the proposed deep-unfolding NN always provides
performance close to the MSER-based GD algorithm.
\begin{table}[t]
	\centering
	\caption{SER for different MIMO system configuration at SNR = $20$ dB.}   \label{table3}
	\begin{small}
		\renewcommand{\arraystretch}{1.4}
		\setlength{\tabcolsep}{0.5mm}{
			\begin{tabular}{c|c|c|c|c}
				\toprule  
				($K$, $R_{r,k}$)& ($1$, $4$) & ($2$, $4$) & ($3$, $2$) & ($4$, $2$) \\
				\hline  
				Hybrid GD & $1.15\!\!\times\!\!10^{-3}$ & $3.64\!\!\times\!\!10^{-3}$ & $6.83\!\!\times\!\!10^{-3}$ & $1.56\!\!\times\!\!10^{-2}$ \\
				\hline  
				Deep-unfolding NN & $98.43$$\%$ & $97.85$$\%$ & $97.07$$\%$ & $95.86$$\%$   \\
				\hline  
				Benchmark CNN & $73.14$$\%$ & $70.48$$\%$ & $68.52$$\%$ & $65.02$$\%$     \\
				\bottomrule  
			\end{tabular}}
	\end{small}
\end{table}
Table \ref{table4} shows the SER performance of the NN schemes versus
the number of training data samples for SNR = $20$ dB. We observe that
the deep-unfolding NN needs much fewer training data samples compared
to the benchmark CNN since its structure is developed based on that of
the MSER-based GD algorithm.
\begin{table*}[t]
	\centering
	\caption{SER versus the number of training samples for SNR = $20$ dB.}   \label{table4}
	\begin{small}
			\renewcommand{\arraystretch}{0.9}
			\setlength{\tabcolsep}{1.5mm}{
				\begin{tabular}{ccccccccc}
				\toprule  
				Training samples & $5000$ & $10000$ & $15000$ & $20000$ & $25000$ & $30000$ & $35000$ & $40000$ \\
				\midrule  
				Benchmark CNN & $66.03$$\%$ & $70.48$$\%$ & $72.55$$\%$ & $73.61$$\%$ & $74.13$$\%$ & $74.21$$\%$ & $74.36$$\%$ & $74.36$$\%$ \\
				\bottomrule  \toprule  
				Training samples & $100$ & $200$ & $300$ & $400$ & $500$ & $600$ & $700$ & $800$   \\
				\midrule  
				Deep-unfolding NN & $92.83$$\%$ & $95.04$$\%$ & $96.21$$\%$ & $97.37$$\%$ & $97.85$$\%$ & $97.94$$\%$ & $98.02$$\%$ & $98.02$$\%$ \\
				\bottomrule  
			\end{tabular}}
	\end{small}
\end{table*}

\begin{figure}[t]
	\centering
	\includegraphics[width=0.35\textwidth]{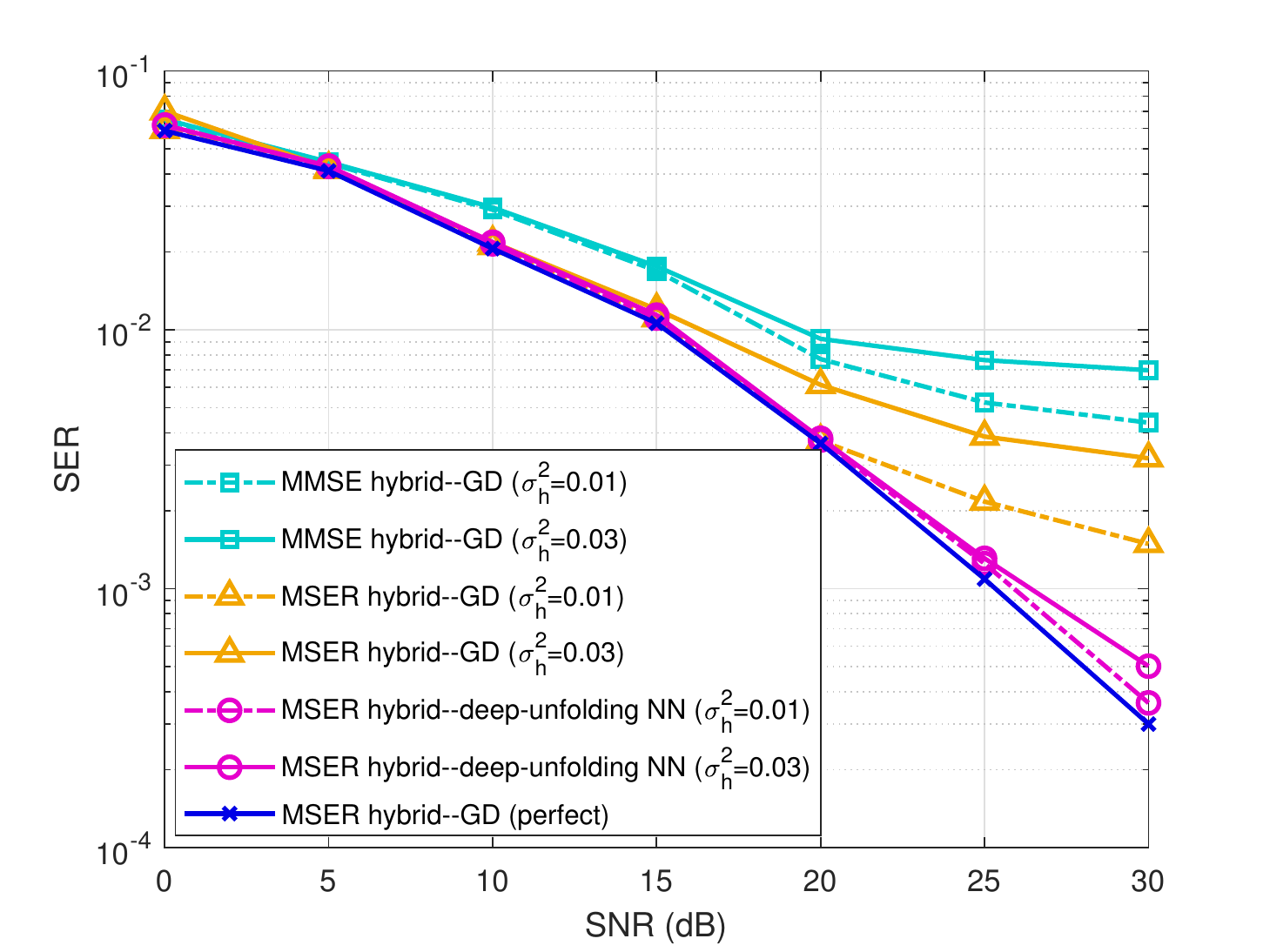}
	\caption{SER performance of different schemes versus SNR in the presence of imperfect CSI (SNR = $20$ dB). }
	\label{result5}
\end{figure}

Fig. \ref{result5} illustrates the SER performance versus SNR of the
various hybrid AD MIMO transceiver designs in the presence of
imperfect CSI.  The channel estimation errors are represented by
$\mathbf{H}_k=\mathbf{\bar{H}}_k + \sigma_{h} \Delta \mathbf{K}_k$,
$\forall k \in \mathcal{K}$, where $\mathbf{H}_k$ is the true
(synthesized) channel matrix, $\mathbf{\bar{H}}_k$ denotes its
estimates, and the term $\sigma_h \Delta \mathbf{K}_k$ is the
estimation error. In this latter term, $\Delta \mathbf{K}_k$ is a
random matrix with zero-mean, unit variance uncorrelated elements
following a circular complex Gaussian distribution, while $\sigma_h^2$
provides the estimation error variance, assumed identical $\forall k$
for simplicity.  We can see that the SER performance degrades with the
error variance $\sigma_{h}^2$. Moreover, the best performance is
achieved by the proposed deep-unfolding NN, followed by the MSER-based
GD beamforming algorithm and the MMSE-based GD beamforming
algorithm. These results show that the proposed deep-unfolding NN has
stronger robustness against the channel uncertainties compared to the
other schemes.

\begin{figure}[t]
	\centering
	\includegraphics[width=0.35\textwidth]{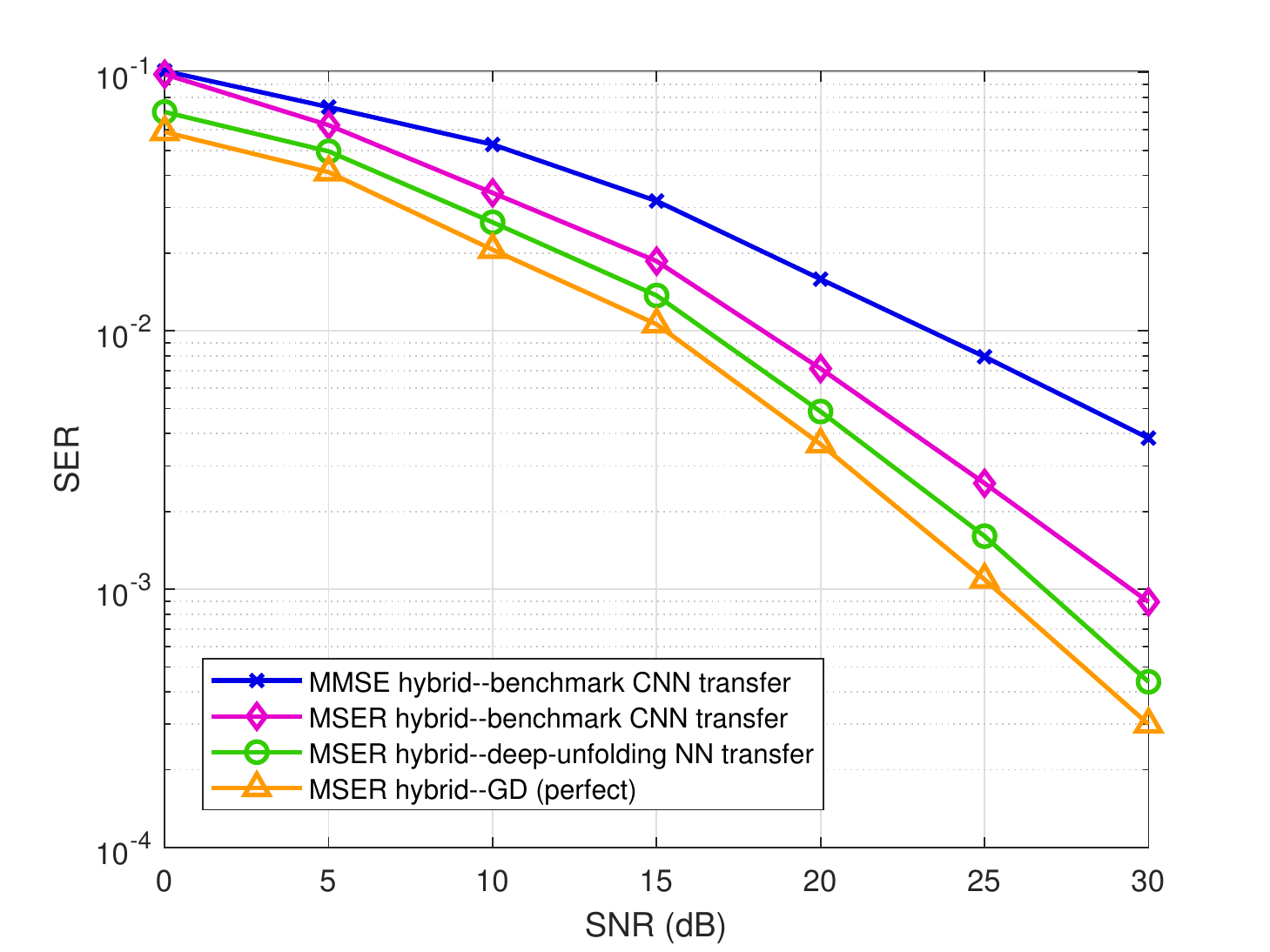}
	\caption{{
			SER performance of different schemes versus SNR under transfer conditions.} }
	\label{transfer_QPSK}
\end{figure}

Fig. \ref{transfer_QPSK} illustrates the transfer ability of the proposed schemes, by plotting their SER performance versus SNR for different channel scenarios. Specifically, in this experiment, we first train a network with data based on the considered mmWave channel characteristics, and then transfer the model to the Gaussian channel model. When retraining the model, the parameters of the first $10$ layers are fixed and only the parameters of the remaining layers are trainable. From the result, we can see that the gap between the transferred deep-unfolding NN and the MSER-based hybrid GD algorithm with perfect CSI is much smaller compared to that of the transferred benchmark CNNs under the MSER and MMSE criteria. It shows that the proposed deep-unfolding NN has a better performance in transfer learning as it can significantly apply the knowledge gained from an older scenario and quickly adapt to a new one with less training time.

Table~\ref{table5} presents the computational complexity of the
proposed schemes for different system configurations, as measured by
the CPU running time of the training and testing stages. From the
results, we can see that the CPU time of both training and testing
stages increases with $N_t$ and $K$. Clearly, the proposed
deep-unfolding NN requires less training time compared to the CNN,
which proves that the former structure can effectively accelerate
convergence and reduce the training overhead. Since the deep-unfolding
NN requires a much smaller number of trainable parameters and the
gradients are computed in close-form in the BP, this makes it more
efficient compared to the CNN which employs the platform
``tensorflow". Moreover, the gap in CPU time between the CNN and the
deep-unfolding NN increases with $N_t$ and $K$, where the training
time is much larger than the testing time. The reason is that there
are more complex computations in the training stage but only the FP
process in the testing stage.
\begin{table*}[t]
	\centering
	\caption{CPU running time of the proposed algorithm.}   \label{table5}
	\begin{small}
		\renewcommand{\arraystretch}{1}
		\setlength{\tabcolsep}{1.5mm}{
			\begin{tabular}{c|cc|cc}
				\toprule  
				\multirow{2}{*}{($N_t$, $K$, $R_r$)} & \multicolumn{2}{c|}{CPU time of training stage (min)} & \multicolumn{2}{c}{CPU time of testing stage (s)}\\ 
				& Deep-unfolding NN & Benchmark CNN & Deep-unfolding NN & Benchmark CNN \\
				\hline  
				($64$, $2$, $4$) & $5.35$ & $8.73$ & $0.01$ & $0.01$ \\
				\hline  
				($64$, $3$, $2$) & $12.46$ & $21.58$ & $0.01$ & $0.1$ \\
				\hline  
				($128$, $4$, $8$) & $84.45$ & $130.79$ & $0.07$ & $0.11$ \\
				\hline  
				($128$, $5$, $6$) & $95.73$ & $200.46$ & $0.10$ & $0.13$ \\
				\hline  
				($128$, $6$, $5$) & $109.48$ & $242.35$ & $0.14$ & $0.17$ \\
				\hline  
				($128$, $7$, $4$) & $129.63$ & $295.43$ & $0.17$ & $0.22$ \\
				\hline  
				($128$, $8$, $4$) & $162.86$ & $360.15$ & $0.24$ & $0.28$ \\
				\bottomrule  
			\end{tabular}}
	\end{small}
\end{table*}

Table \ref{table6} aims to show the generalization ability of the
proposed deep-unfolding NN. To this end, we first train a large-scale
network with $N_t = 128$, $R_t = 32$, $N_r = 16$, $R_r = 4$, $K = 8$,
and $D_k = 3$ and then implement it to test the performance of a MIMO
system with smaller number of users $K$ and transmit antennas
$N_t$. From the results, we can see that the performance loss of
applying this large-scale network to test the scenarios with smaller
$K$ and $N_t$ is slight, which indicates the strong generalization
ability of the proposed deep-unfolding NN.
\begin{table}[t]
	\centering
	\caption{Generalization ability of the proposed deep-unfolding NN.}   \label{table6}
	\begin{small}
		\renewcommand{\arraystretch}{1.2}
		\setlength{\tabcolsep}{3mm}{
			\begin{tabular}{c|cccc}
				\toprule  
				\diagbox{$N_t$}{$K$} & $8$ & $6$ & $4$ & $2$ \\
				\hline  
				$128$ & $98.04$$\%$ & $98.45$$\%$ & $98.93$$\%$ & $99.37$$\%$ \\
				\hline  
				$64$ & \textemdash & \textemdash & $96.94$$\%$ & $97.52$$\%$ \\
				\bottomrule  
			\end{tabular}}
	\end{small}
\end{table}

\vspace{-4mm}
\subsection{SER Performance for 16-QAM Signals}

\begin{figure}[t]
	\centering
	\includegraphics[width=0.35\textwidth]{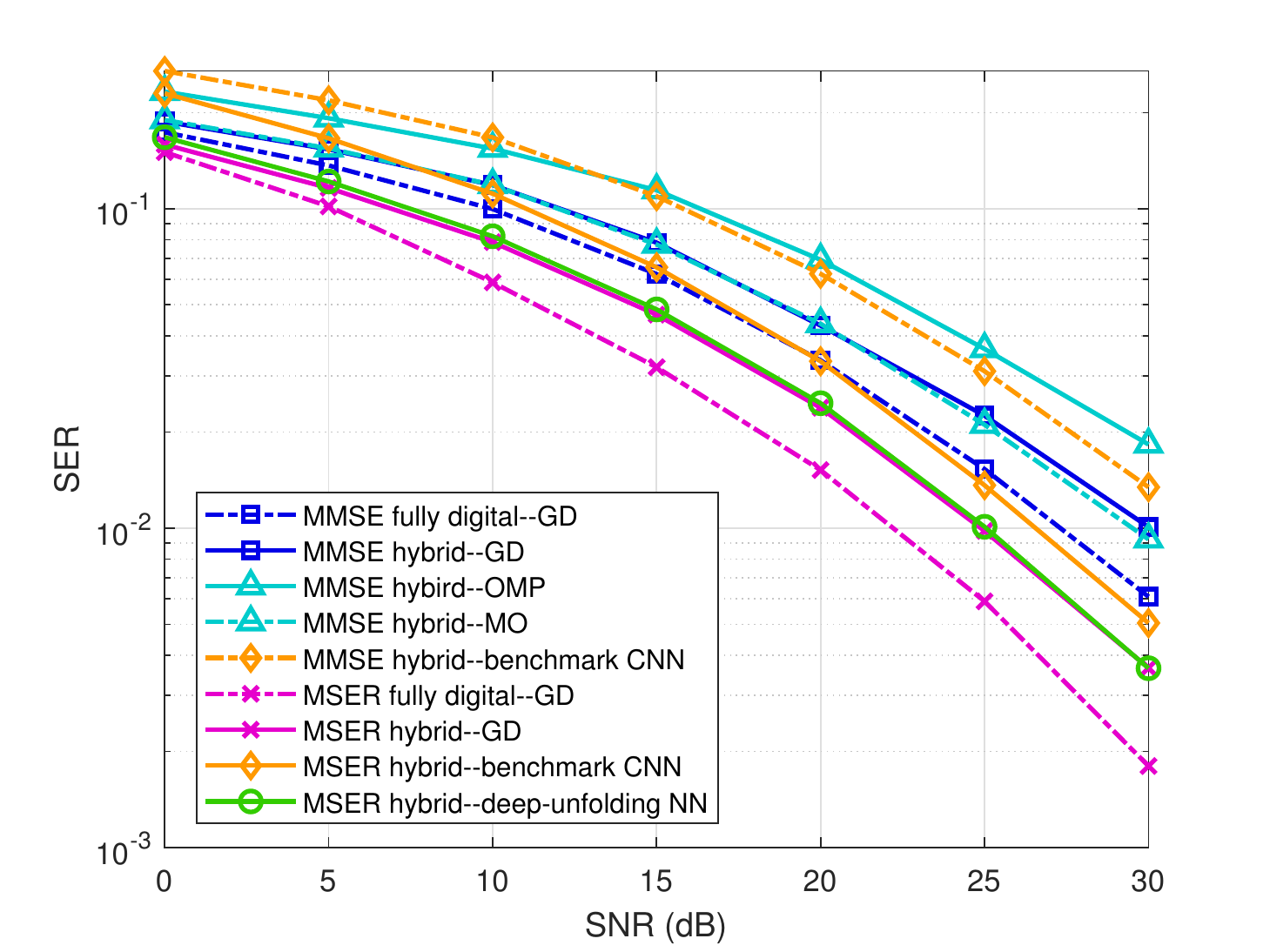}
	\caption{{
			SER performance of the different schemes versus SNR for 16-QAM signals.}}
	\label{result6}
\end{figure}

Fig. \ref{result6} presents the SER performance of the different
schemes versus SNR in the case of 16-QAM modulation. Similar to the
QPSK modulation, the MSER-based algorithms achieve better performance
than the MMSE-based algorithms. Meanwhile, the performance of the
proposed deep-unfolding NN approaches that of the MSER-based hybrid GD
beamforming algorithm although the performance gap between these two
schemes decreases with SNR. In addition, we observe that the proposed
deep-unfolding NN significantly outperforms the benchmark CNN at lower
SNR. Table \ref{table7} provides the SER values along with the number
of the iterations/layers for SNR $= 25$ dB. It demonstrates the
ability of the deep-unfolding NN to achieve nearly the same
performance as the hybrid GD method while significantly reducing
computational complexity. The above results show that the proposed
deep-unfolding NN still works well for other higher-order modulation
schemes.

\begin{table}[t]
	\centering
	\caption{SER versus the iterations/layers for SNR = $25$ dB.}   \label{table7}
	\begin{small}
		\renewcommand{\arraystretch}{1.4}
		\setlength{\tabcolsep}{0.4mm}{
			\begin{tabular}{c|c|c|c}
				\toprule  
				\multirow{2}{*}{} & Hybrid GD & Benchmark CNN & Deep-unfolding NN \\
				& $500$ iterations & $17$ layers & $15$ layers \\
				\hline  
				\ SNR = $25$ dB  & $9.8 \times 10^{-3}$ & $1.36 \times 10^{-2}$ & $1.01 \times 10^{-2}$   \\
				\bottomrule  
			\end{tabular}}
	\end{small}
\end{table}

\begin{figure}[t]
	\centering
	\includegraphics[width=0.35\textwidth]{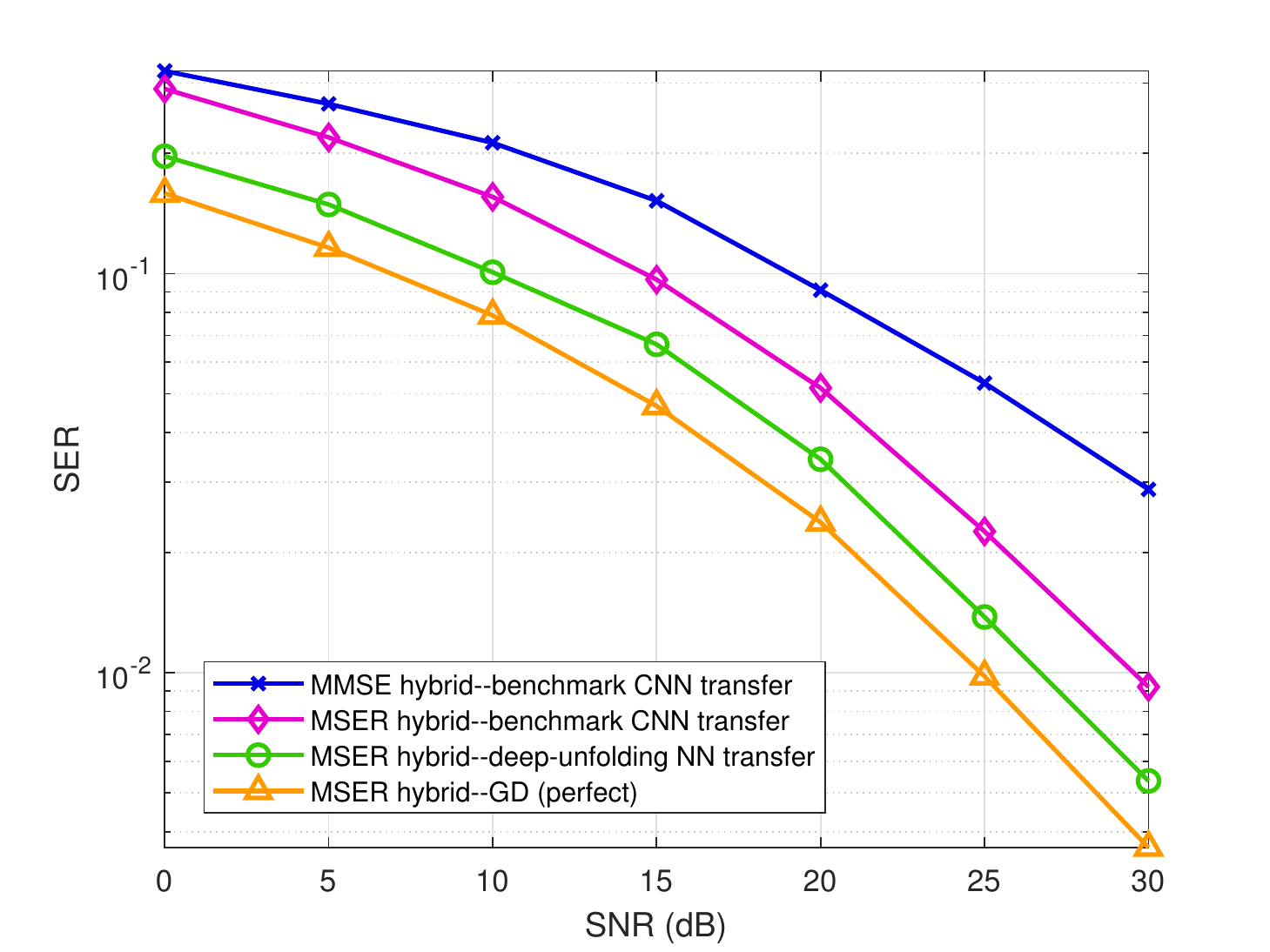}
	\caption{{
			SER performance of different schemes versus SNR under transfer conditions.} }
	\label{transfer_QAM}
\end{figure}

Fig. \ref{transfer_QAM} illustrates the transfer ability of the proposed shemes for 16-QAM signals. We can see that the gap between the deep-unfolding NN and the MSER-based hybrid GD algorithm is much smaller compared to that of the benchmark CNNs for changed channel characteristics. The results show that the proposed deep-unfolding NN has a better performance in transfer learning compared to the benchmark CNNs.

\vspace{-3mm}
\section{Conclusion}
\label{Section8:conclusion}

In this work, we addressed the problem of hybrid AD transceiver design
for massive-MIMO systems based on the MSER criterion.  Following the
mathematical formulation of the problem as a constrained optimization,
we first developed a MSER-based GD iterative algorithm to find
stationary points. We then unfolded the GD iterative algorithm into a
multi-layer structure and proposed a deep-unfolding NN, where a set of
trainable parameters are introduced to reduce the computational
complexity and improve the system performance. 
For the purpose of training,
we obtained the relationship linking the gradients between two
adjacent layers based on the GCR. The deep-unfolding NN was proposed
for both QPSK and $M$-QAM signal constellations and its convergence was investigated through theoretical analysis; besides, we
developed a black-box NN based on a recently proposed approach, for use  as a benchmark. Simulation results showed
that the proposed deep-unfolding NN provides better SER performance
compared to the black-box CNN and approaches the performance of the
MSER-based GD iterative algorithm with much lower
complexity. Extensions of the deep-unfolding schemes developed in this
paper to nonlinear modulation techniques or other types of beamforming
algorithms are interesting avenues for future work.
\vspace{-0mm}

\vspace{-2mm}
\begin{appendices}

\section{GCR in Matrix Form}
\label{appendixA}
Based on \cite{IAIDNN}, the general form of the optimization problem can be formulated as \vspace{-1mm}
\begin{equation}
	\min \limits_{\mathbf{X}} \ g(\mathbf{X}; \mathbf{Z}) \quad \st \ \mathbf{X} \in \mathcal{S},  \vspace{-1mm}
\end{equation}
where $g:\mathbb{C}^{m \times n} \rightarrow \mathbb{R}$ denotes a continuous objective function, $\mathbf{X} \in \mathbb{C}^{m \times n}$ denotes the variable, $\mathcal{S}$ is the feasible region, and $\mathbf{Z}$ is the random parameter in the problem.
To solve the problem, an iterative algorithm is developed as \vspace{-1mm}
\begin{equation}
	\mathbf{X}^{t+1} = G_t(\mathbf{X}^t; \mathbf{Z}),  \vspace{-1mm}
\end{equation}
where function $G_t(\cdot)$ maps the variable $\mathbf{X}^t$ to the variable $\mathbf{X}^{t+1}$ in the $t$-th iteration.
To reduce the computational complexity, trainable parameter $\bm{\lambda}$ is introduced. Since $\mathbf{Z}$ is a random variable, we need to take the expectation of $\mathbf{Z}$ and the problem is  transformed into \vspace{-1mm}
\begin{equation}
	\min \limits_{\mathbf{X}} \ \mathbb{E}_{\mathbf{Z}} \{g(\mathbf{X}; \bm{\lambda}, \mathbf{Z}) \} \quad \st \ \mathbf{X} \in \mathcal{S}.  \vspace{-1mm}
\end{equation}

A deep-unfolding NN can be developed for the above problem as follows, \vspace{-1mm}
\begin{equation}
	\mathbf{X}^{l+1} = \mathcal{G}_l(\mathbf{X}^l; \bm{\lambda}^l, \mathbf{Z}), \vspace{-2mm}
\end{equation}
where $\mathcal{G}_l(\cdot)$ denotes the update function of the NN in the $l$-th layer, $\mathbf{X}^{l}$ and $\mathbf{X}^{l+1}$ denote the input and output of the $l$-th layer, respectively, $\mathbf{Z}$ represents the fixed parameter or input of the NN, and $\bm{\lambda}^l$ is the trainable parameter in the $l$-th layer. For such a NN, in order to train the parameter $\bm{\lambda}^l$, we need to derive the relationship between the gradients of adjacent layers and then calculate the gradient w.r.t. $\bm{\lambda}^l$. To apply the GCR which leads to an expression of the following type \vspace{-0.5mm}
\begin{equation}
	{\rm Tr} \left(\mathbf{G}^{l+1} \, d\mathbf{X}^{l+1} \right)  =  {\rm Tr} \left(\mathbf{G}^{l+1} \circ \mathcal{J}(\mathbf{X}^l; \bm{\lambda}^l, \mathbf{Z}) \, d\mathbf{X}^l \right),  \vspace{-1mm}
\end{equation}
where $\mathbf{G}^{l+1}$ and $\mathbf{G}^l$ are the gradients w.r.t. $\mathbf{X}^{l+1}$ and $\mathbf{X}^l$, respectively, $\mathcal{J}(\mathbf{X}^l; \bm{\lambda}^l, \mathbf{Z})$ denotes some matrix functions of $\mathbf{X}^l$, $\bm{\lambda}^l$, and $\mathbf{Z}$, which are related to the update function $\mathcal{G}_l(\cdot)$. Then, we obtain the desired gradient relationship as \vspace{-1mm}
\begin{equation}
	\mathbf{G}^l = \mathbf{G}^{l+1} \circ \mathcal{J}(\mathbf{X}^l; \bm{\lambda}^l, \mathbf{Z}). \vspace{-1mm}
\end{equation}
In this work, when $\mathbf{X}$ denotes $\mathbf{P}_k$, we find that $\mathcal{J}(\cdot) = \mathcal{J}_1(\cdot) \circ \mathcal{J}_2(\cdot)$, where $\mathcal{J}_1(\cdot) = \mathbf{E}^{\perp} -  b_k^H B  (\bm{\alpha}_{P_k}^{l})^T (\mathbf{F}^{l})^H \mathbf{H}_k^H$
$(\mathbf{U}_k^{l})^H \mathbf{W}_k^{l}$ and $\mathcal{J}_2(\cdot) = \mathbf{E}$, where $\mathbf{E}$ is given in Section \ref{deep-unfolding-back}.

\vspace{-2mm}
\section{Gradients of Trainable Parameters}
\label{appendixB}
According to the update rules in \eqref{update_p_new}--\eqref{update_F_new}, the gradients w.r.t. the introduced trainable parameters in each layer are acquired as follows, where the index of iteration $t$ is omitted for simplicity. \vspace{-1mm}
	\begin{equation}             \label{G_parameters}
	\begin{split}
		&\!\! \nabla_{\bm{\alpha}_{P_k}} \tilde{\mathcal{P}}_e^{l} = -(\mathbf{G}_{P_k}^{l+1})^H \circ (\nabla_{\mathbf{P}_k^{H}} \tilde{\mathcal{P}}_e^{l})^H,
		\nabla_{\mathbf{O}_{P_k}} \tilde{\mathcal{P}}_e^{l} = (\mathbf{G}_{P_k}^{l+1})^H,
		\\
		& \!\!\nabla_{\bm{\alpha}_{\theta_F}} \tilde{\mathcal{P}}_e^{l} = -(\mathbf{G}_{\!F}^{\!l\!+\!1})^{\!H} \!\circ\! j \mathbf{F}^* \!\circ\! (\nabla_{\!\mathbf{F}} \tilde{\mathcal{P}}_{\!\!e}^{l})^{\!H} \!+\! (\mathbf{G}_{\!F^{\!*}}^{\!l\!+\!1})^{\!H} \!\circ\! j \mathbf{F} \!\circ\! (\nabla_{\!\mathbf{F}^{*}} \!\tilde{\mathcal{P}}_{\!\!e}^{l})^{\!H},
		\\
		&\!\! \nabla_{\mathbf{O}_{\theta_F}} \tilde{\mathcal{P}}_e^{l} = (\mathbf{G}_F^{l+1})^H \circ j \mathbf{F}^* - (\mathbf{G}_{F^*}^{l+1})^H \circ j \mathbf{F},
		\\
		& \!\nabla_{\rho_{P_k}} \tilde{\mathcal{P}}_e^{l} =  \frac{1} {J \sqrt{2\pi} (\rho_{P_k}^{l})^2 }  \sum \limits_{{j}=1}^J  \sum \limits_{i=1}^{D_k} e^{-\frac{|\bar{b}^{{q_j}}_{i,k}|^2} {2(\rho_{P_k}^{l})^2}} \! \Big(1-\frac{|\bar{b}^{{q_j}}_{i,k}|^2} {2(\rho_{P_k}^{l})^2} \Big) \mathbf{G}_{P_k}^{l+1}
		\\
		&\qquad \quad \circ  (\bm{\alpha}_{P_k}^{l})^T  \mathbf{F}^H \mathbf{H}_k^H \mathbf{U}_k^H \mathbf{w}_{i,k}.
	\end{split}
\end{equation}
\end{appendices}

\vspace{-6mm}
\bibliography{references}


\begin{spacing}{0.93}

\end{spacing}

\end{document}